\documentclass[fleqn, usenatbib]{mnras}

\usepackage{amssymb,amsmath}
\usepackage{subfig}
\usepackage{graphicx}
\usepackage{color}
\usepackage{commath}
\usepackage{listings}
\usepackage{wrapfig}
\usepackage{float}
\usepackage{tikz}
\usetikzlibrary{arrows,decorations.pathmorphing,backgrounds,positioning,fit,petri}
\usetikzlibrary{decorations.markings}
\usetikzlibrary{arrows.meta}
\usepackage{mathtools}

\usepackage{eso-pic}

\AddToShipoutPictureBG*{%
  \AtPageUpperLeft{%
    \hspace{0.75\paperwidth}%
    \raisebox{-3.5\baselineskip}{%
      \makebox[0pt][l]{\textnormal{DES 2018-0352}}
}}}%

\AddToShipoutPictureBG*{%
  \AtPageUpperLeft{%
    \hspace{0.75\paperwidth}%
    \raisebox{-4.5\baselineskip}{%
      \makebox[0pt][l]{\textnormal{FERMILAB-PUB-19-227-AE}}
}}}%

\tikzset{bigarrow/.style={decoration={markings,mark=at position 1 with {\arrow[scale=40]{>}}},postaction={decorate}}}

\begin{document}
\lstset{frame=tb,
  	language=Matlab,
  	aboveskip=3mm,
  	belowskip=3mm,
  	showstringspaces=false,
  	columns=flexible,
  	basicstyle={\small\ttfamily},
  	numbers=none,
  	numberstyle=\tiny\color{gray},
 	keywordstyle=\color{blue},
	commentstyle=\color{dkgreen},
  	stringstyle=\color{mauve},
  	breaklines=true,
  	breakatwhitespace=true
  	tabsize=3
}

\title[DES Y1 mass maps via forward fitting in harmonic space]{Dark Energy Survey Year 1 Results: Wide field mass maps via forward fitting in harmonic space}

\author[DES Collaboration]{
\parbox{\textwidth}{
\Large
B.~Mawdsley,$^{1}$
D.~Bacon,$^{1}$
C.~Chang,$^{2,3}$
P.~Melchior,$^{4}$
E.~Rozo,$^{5}$
S.~Seitz,$^{6,7}$
N.~Jeffrey,$^{8}$
M.~Gatti,$^{9}$
E.~Gaztanaga,$^{10,11}$
D.~Gruen,$^{12,13,14}$
W.~G.~Hartley,$^{8,15}$
B.~Hoyle,$^{6,7}$
S.~Samuroff,$^{16}$
E.~Sheldon,$^{17}$
M.~A.~Troxel,$^{18}$
J.~Zuntz,$^{19}$
T.~M.~C.~Abbott,$^{20}$
J.~Annis,$^{21}$
E.~Bertin,$^{22,23}$
S.~L.~Bridle,$^{24}$
D.~Brooks,$^{8}$
E.~Buckley-Geer,$^{21}$
D.~L.~Burke,$^{13,14}$
A.~Carnero~Rosell,$^{25,26}$
M.~Carrasco~Kind,$^{27,28}$
J.~Carretero,$^{9}$
L.~N.~da Costa,$^{26,29}$
J.~De~Vicente,$^{25}$
S.~Desai,$^{30}$
H.~T.~Diehl,$^{21}$
P.~Doel,$^{8}$
A.~E.~Evrard,$^{31,32}$
B.~Flaugher,$^{21}$
P.~Fosalba,$^{10,11}$
J.~Frieman,$^{21,3}$
J.~Garc\'ia-Bellido,$^{33}$
D.~W.~Gerdes,$^{31,32}$
R.~A.~Gruendl,$^{27,28}$
J.~Gschwend,$^{26,29}$
G.~Gutierrez,$^{21}$
D.~L.~Hollowood,$^{34}$
K.~Honscheid,$^{35,36}$
D.~J.~James,$^{37}$
M.~Jarvis,$^{38}$
T.~Jeltema,$^{34}$
K.~Kuehn,$^{39}$
N.~Kuropatkin,$^{21}$
M.~Lima,$^{40,26}$
M.~A.~G.~Maia,$^{26,29}$
J.~L.~Marshall,$^{41}$
R.~Miquel,$^{42,9}$
A.~A.~Plazas,$^{4}$
A.~Roodman,$^{13,14}$
E.~Sanchez,$^{25}$
V.~Scarpine,$^{21}$
S.~Serrano,$^{10,11}$
I.~Sevilla-Noarbe,$^{25}$
M.~Smith,$^{43}$
R.~C.~Smith,$^{20}$
F.~Sobreira,$^{44,26}$
E.~Suchyta,$^{45}$
M.~E.~C.~Swanson,$^{28}$
G.~Tarle,$^{32}$
D.~L.~Tucker,$^{21}$
V.~Vikram,$^{46}$
and A.~R.~Walker$^{20}$
\begin{center} (DES Collaboration) \end{center}
}}
\vspace{0.2cm}

\maketitle
\begin{abstract}
We present new wide-field weak lensing mass maps for the Year 1 Dark Energy Survey data, generated via a forward fitting approach. This method of producing maps does not impose any prior constraints on the mass distribution to be reconstructed. The technique is found to improve the map reconstruction on the edges of the field compared to the conventional Kaiser-Squires method, which applies a direct inversion on the data; our approach is in good agreement with the previous direct approach in the central regions of the footprint. The mapping technique is assessed and verified with tests on simulations; together with the Kaiser-Squires method, the technique is then applied to data from the Dark Energy Survey Year 1 data and the differences between the two methods are compared. We also produce the first DES measurements of the convergence Minkowski functionals and compare them to those measured in simulations. \\ 
key words:gravitational lensing: weak, large-scale structure of Universe, dark matter
\par 
\vspace{1cm}
\end{abstract}

\section{Introduction}
The presence of a gravitational potential influences light trajectories so that, to an observer, the apparent position and shape of objects is altered via an effect known as \textit{gravitational lensing}. As this gravitational potential is related to the matter distribution, there is a relationship between the strength of the lensing effect and the density contrast from which it arises. The arrangement of a source of light, a mass distribution acting as a lens, and an observer can produce a range of magnitudes of lensing effects. This paper pertains to weak lensing, which produces small  $\lesssim$1\% changes in the source objects'  ellipticities on an image. Measurements of galaxy ellipticities can therefore be used to estimate lensing \textit{shear}, and hence constrain the integrated gravitational potential and the integrated matter distribution between the observer and a lensed galaxy. Shear can be related to the matter surface density, or \textit{convergence} distribution $\kappa$, through a technique known as the Kaiser-Squires inversion \citep{KS-1993ApJ...404..441K} to produce large scale maps of the integrated mass distribution. 
\par

As the perturbations to the underlying metric of the Universe from matter concentrations are solely caused by gravitation, lensing is only sensitive to the total density contrast and not to any further properties that the matter may have. Lensing is therefore a probe of the dark Universe as well as luminous matter. This is in marked contrast to using other probes, such as galaxy clustering, where luminous matter is a proxy for the location of dark matter and we must introduce a bias parameter. Lensing, and the matter distributions inferred from it, are  powerful tools to constrain cosmological parameters - and this is not just true of the two-point statistics of shear or convergence; the $\kappa$ maps contain further information on the phase of the matter distribution.  Additional non-Gaussian information in the fields has been examined through peak statistics of the convergence field \citep{Peaks3doi:10.1111/j.1365-2966.2009.15948.x, PeaksPhysRevD.81.043519, Peaks4PhysRevD.84.043529,Peaksdoi:10.1093/mnras/stw2070}   Minkowski Functionals \citep{MF2PhysRevD.88.123002, PetriMF:2015ura, DMMF2012MNRAS.419..536M}, 3 point statistics \citep{Dod3ptPhysRevD.72.083001} and through examination of the full distribution of values in the convergence map, or Probability Density Function \citep{PDF12015MNRAS.448.1389C, PDF2_Patton:2016umg}. Maps are also useful for cross-correlating with other observations, for example by constraining bias as in \citet{ChangBias_doi:10.1093/mnras/stw861}. 

\par
With the next generation of surveys covering significant fractions of the sky and measuring large numbers of galaxy shapes to high precision, weak lensing promises to be a leading cosmological probe in coming years, and will map the large-scale structure of the Universe to an unprecedented scale. Wide field mass maps have been produced using weak lensing measurements in a number of recent surveys: the Canada-France Hawaii Telescope Lensing Survey \citep{CFHT-doi:10.1093}, the Cosmic Evolution Survey  \citep{MasseyCosmos2007Natur.445..286M} and the Kilo-Degree Survey \citep{KIDS_refId0}. The Dark Energy Survey \citep{DES_summary_doi:10.1142/S0217751X05025917} has produced mass maps through the Kaiser-Squires inversion, on an approximately flat field in their Science Verification data  \citep{Vikram_mass:2015leg, Chang:2015odg} and on the sphere for the larger Year 1 footprint \citep{Chang:2017kmv}. The Hyper Suprime-Cam survey \citep{HSCa_2017arXiv170208449A, HSCb_2017arXiv170405858A} survey has produced mass maps \citep{Mandelbaum:2017dvy} over a 167 degrees$^{2}$ area of the sky, and tomographically across several redshift bins \citep{Oguri:2017vrv}, resolving structure down to finer resolution due to its high galaxy density, allowing for smaller scale shear measurements.  \par

In this paper, we produce updated mass maps using the Dark Energy Survey Year 1 data. The previously produced maps were made through directly transforming shear fields into convergence maps, which introduces errors around the edges of the survey footprint; our alternative method avoids directly transforming data and prevents the introduction of these edge effects, by fitting hypothesis full-sky maps to the data. We examine how this method compares to the previous approach, and analyse its potential for further applications to wide field mass mapping. \par
This paper is arranged into 5 sections. Section \ref{sec:lens_maths} describes the weak lensing formalism; then we introduce the forward fitting approach in section \ref{sec:method}. Section \ref{sec:tests} provides details of the tests undertaken to examine the resultant maps, and Section \ref{sec:applying_routine} presents the maps produced using the DES Y1 data. We conclude in Section \ref{sec:conclusion}.

\section{Lensing Formalism}
\label{sec:lens_maths}
Here we briefly describe the relevant weak lensing formalism for our work. A comprehensive review of weak lensing can be found in \cite{Bartelmann:1999yn}; in the following section we shall follow the spherical harmonics approach described in
\cite{Castro-PhysRevD.72.023516}. The large observed area of modern surveys mean that a full sky treatment is required, which is achieved with techniques commonly used in CMB analyses \citep{Leistedt_doi:10.1093/mnras/stw3176, HEavens03_doi:10.1046/j.1365-8711.2003.06780.x, Kitch14_doi:10.1093/mnras/stu934}. \par 

We can define a \textit{lensing potential} $\phi$ at a given spatial coordinate $\textbf{r} = (r, \theta, \psi) $ by  
\begin{equation}
\phi(r, \theta, \psi ) = \frac{2}{c^2} \int_0^r dr' \frac{f_K(r-r')}{f_K(r)f_K(r')} \Phi(r', \theta, \psi)
,\end{equation}
where $\Phi$ is the Newtonian potential and $f_k$ is a comoving angular diameter distance, taking values of ($\sin r, r ,$ or $ \sinh r$) for a Universe with curvature described by $k=1$  (closed), $0$  (flat) or $ -1 $ (open), \cite{Castro-PhysRevD.72.023516}. The coordinate $r$ is a radial distance and $(\theta, \phi)$ refer to angular positions on the sky. This potential can be related to the matter density through Poisson's equation
\begin{equation}
\nabla_r^2 \Phi(\textbf{r}) = \frac{3\Omega_m H_0^2}{2a(t)}\delta(\textbf{r})
,\end{equation}
where $\Omega_m$ is the present day total matter density parameter, $H_0$ is the Hubble constant at the present time, $a(t)$ is the scale factor, and $\delta(\textbf{r})$ is the density contrast at position $\textbf{r}$. \par
We shall consider the case of a scalar field $\phi(\textbf{r})$ in a flat background geometry, which can be transformed into the basis of spherical harmonics and spherical Bessel functions via 
\begin{equation}
\phi_{\ell m}(k)=\sqrt[]{\frac{2}{\pi}} \int \mathrm{d}^3 r \phi(\textbf{r}) k j_{\ell}(kr) Y^{*}_{\ell m}(\theta, \varphi)
\end{equation}
Through the introduction of a geometrical differential operator $\eth (\bar{\eth})$ , which raises (lowers) the spin of the field, $\psi$ can be related to the shear $\gamma$ and convergence $\kappa$ through the following relations:
\begin{equation}
\kappa(\textbf{r})=\frac{1}{4}(\eth \bar{\eth} + \bar{\eth}\eth) \phi(\textbf{r})  
,\end{equation}
\begin{equation}
\gamma(\textbf{r}) = \frac{1}{2} \eth \eth \phi (\textbf{r})
,\end{equation}
where $\gamma(\textbf{r})$ is composed of two orthogonal components
\begin{equation}
\begin{aligned}
\gamma_1 (\textbf{r}) &= \frac{1}{4}(\eth \eth + \bar{\eth}  \bar{\eth} ) \phi(\textbf{r}) \\
\gamma_2 (\textbf{r}) &= -\frac{i}{4}(\eth \eth - \bar{\eth}  \bar{\eth} ) \phi(\textbf{r})
\end{aligned}
\end{equation}
Furthermore, as shear is a spin-2 field, it will decompose into spin-2 weight spherical harmonics ($\prescript{}{2} \gamma_{\ell m}$), known as E and B modes, that are free of curl and divergence respectively. In the single thin lens plane case, it can be shown that the lensing information is contained within the E mode coefficients \citep{Castro-PhysRevD.72.023516}, and that the B mode coefficients should only be non-zero in the presence of noise. Multiple lenses can give rise to a small B mode, but these will be negligible given our signal-to-noise. Because of the $\eth$ operator rules for spherical harmonics, the equations relating coefficients for $\phi, \kappa $, $\gamma$ and lensing deflection $\alpha$ are:
\begin{equation}
\prescript{}{2}\gamma_{\ell m} (k) =- A_{E, \ell m}= \frac{1}{2} \sqrt[]{\frac{(\ell+2)!}{(\ell-2)!}}\phi_{\ell m}(k)
\label{eq;Egam_rel}
,\end{equation}
\begin{equation}
\kappa_{\ell m}(k)= -\frac{\ell(\ell+1)}{2}\phi_{\ell m}(k)
\label{eq:kap_ell}
,\end{equation}
\begin{equation}
\alpha_{\ell m}= \sqrt[]{\ell(\ell+1)} \phi_{\ell m}(k)
\label{eq:alpha_ell}
,\end{equation}
 where $\phi_{\ell m}(k)$ are the coefficients of the lensing potential in spherical harmonics. Using these quantities in harmonic space, we can transform an observed sky shear signal to maps of the quantities $\kappa, \alpha $ and $\phi$.

Shear is estimated using galaxy ellipticities $\epsilon$; by averaging over a significant number of galaxies and in the absence of intrinsic alignments, ellipticities not due to lensing should average to give a (noisy) mean ellipticity of zero;  any remaining signal is due to the lensing shear $\gamma$, i.e
\begin{equation}
\epsilon = \gamma + \epsilon_{int} + \epsilon_s
,\end{equation}
where $\epsilon_s$ is the noise associated with estimating a galaxy shape, $\epsilon_{int}$ is the intrinsic shape of the galaxy and $\epsilon$ is the observed distortion. Averaging this estimator for a large number of galaxies will reduce the noise, so it is desirable to have the densest possible background field of lensed galaxies.

\section{Methodology}
\label{sec:method}
 There are two  methods used in this paper - the  Kaiser-Squires reconstruction applied directly to the data, and our forward-fitting method utilising hypothesised full-sky shear fields. Both are expressed in spherical harmonics, and use the same $\gamma$ to $\kappa$ pipeline. However, they do differ through the shear fields used, as the forward fitting model uses hypothesis full-sky shear fields, while the direct inversion uses shear data from the survey footprint. \par
In order to produce maps on the scales covered by DES Y1, a package that utilises the spherical approach is needed. We use the \texttt{healpy} suite which is a python wrapper for \texttt{HEALPIX}\footnote{http://healpix.sf.net/}, software which is designed to handle data on the sphere and initially developed for use with the cosmic microwave background. The spin-2 $\gamma$ observations are analogous to the polarisation Stokes parameters Q and U used in CMB studies. Estimated maps of $\gamma_1$ and $\gamma_2$ are entered as arguments in the function \texttt{map2alm} to produce their spherical harmonics coefficients in the form of the divergence free $B_{\ell m}$ and the curl free $E_{\ell m}$. \par
\subsection{Direct inversion}
\begin{figure*}
\centering
\includegraphics[width=\textwidth]{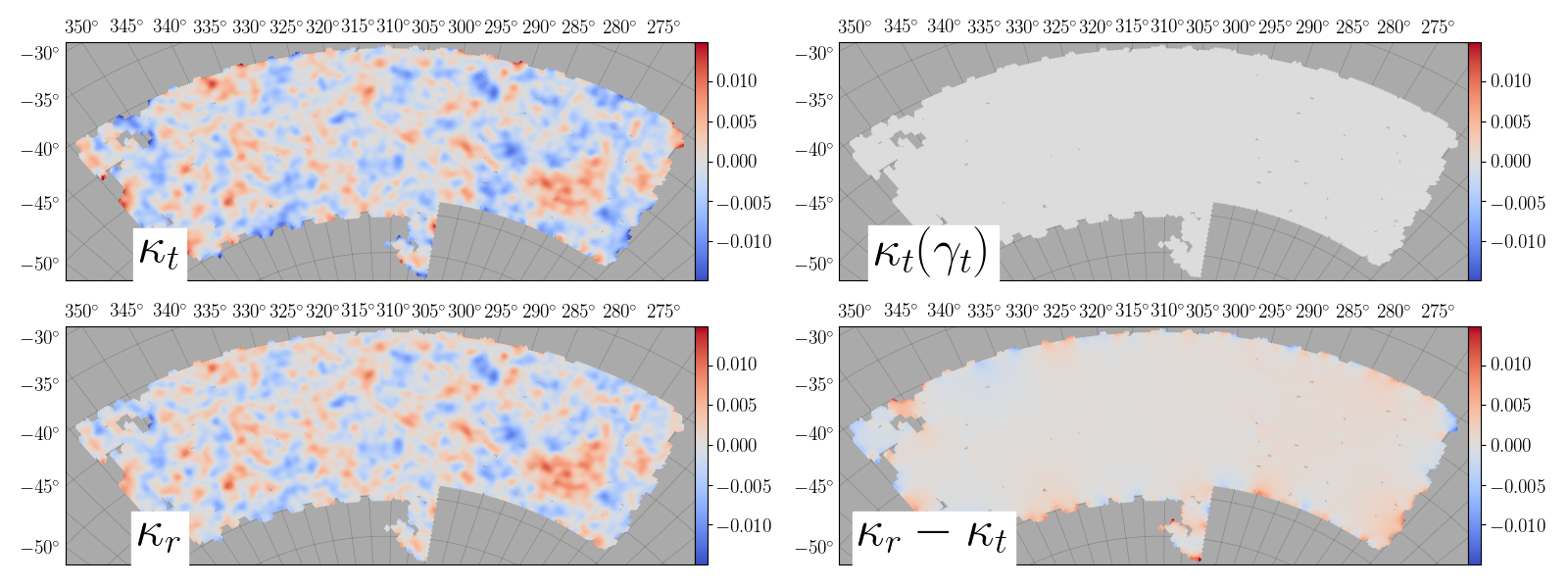}
\caption{Comparison of reconstructions performed using $\kappa_t$, $\gamma_1$ and $\gamma_2$ on maps of \texttt{nside}$=256$, using the Kaiser Squires direct inversion technique. These pixels have a width of $\sim 0.23$ deg and the maps are smoothed with a Gaussian kernel with  $\sigma = 20$ deg. Upper left shows a true $\kappa$ distribution, and upper right shows the residuals for a reconstruction of this field when using the full sky of $\gamma_1$ and $\gamma_2$ values, without shape noise added. Lower left shows the reconstruction of the map using shear values from only the survey region, and the lower right the residuals for this approach. Comparing the two right hand plots shows clearly that considerable residuals can arise in the edge pixels of the survey footprint, even before the addition of shape noise, and the scale of this error can be of a similar size to the $\kappa$ field that we are attempting to reconstruct. 
}
\label{fig:reconcomp}
\end{figure*}
Kaiser-Squires on the sphere uses the E modes found from a spherical transform of a shear field with equation (\ref{eq;Egam_rel}) to produce convergence coefficients; the B mode coefficients  $B_{\ell m} $ serve as a useful null test for possible systematics and noise. Final maps are produced via the reverse transform \texttt{alm2map}, with \textbf{$\kappa$} and \textbf{$\phi$} using the spin-0 case of the transform and \textbf{$\alpha$} the spin-1 case (equations \ref{eq:kap_ell} and \ref{eq:alpha_ell}). The output is a \texttt{healpy} map, pixelised to a chosen resolution, and then smoothed using the \texttt{smoothing} function. Producing maps at a higher resolution reduces the magnitude of the edge effects, and smoothing improves the signal to noise. All maps in this paper are displayed using the Albers Equal Area projection and using \texttt{SKYMAPPER} \footnote{https://github.com/pmelchior/skymapper}.\par
Figure \ref{fig:reconcomp} shows simulated skies reconstructed with the direct inversion.  Using a full sky of known $\gamma$ (right top), the reconstruction from $\gamma$ to $\kappa$ has negligible residuals. However, when a limited shear data area is used, the reconstruction introduces significant errors along the edges of the data footprint (right bottom). These arise due to the assumption that the complete field is sampled, when in reality it is not; unobserved regions of the sky are treated as having zero shear and some bias in the true harmonic coefficients is introduced. These errors are significant fractions of the typical $\kappa$ values, which is apparent when comparing the size of residuals in Figure \ref{fig:reconcomp} to the true $\kappa$ field. This can result in significant contamination of the recovered map. In this paper, we will refer to making maps this way, from a finite survey area, as the \textit{direct inversion} method. 

\subsection{Forward fitting approach}
We have developed a forward fitting approach with the aim of mitigating some of the limitations of the direct inversion. The direct inversion method was initially developed with two assumptions - that the field is uniform in its noise, and that the shear field is completely observed. Creating mass maps with a field that has been observed over a fraction of the sky and with a non-uniform noise distribution will violate these assumptions and therefore introduce errors in the reconstructed map \citep{Seitz_1995A&A...297..287S, Seitz_1998astro.ph..2051S}. The motivation for the forward-fitting technique is therefore to produce maps using the usual relationship between $\gamma$ and $\kappa$ fields, but to do so in such a way that they are made without transforming the limited shear measurements directly into spherical harmonics.
\par
Our technique instead hypothesises a full sky of shear values, and then compares this hypothesis to the observations. Thus, the masked data do not directly enter the transform and the edge effects introduced by the direct inversion are avoided. 
Comparing a region of the hypothesised sky $\gamma$ values to the observed $\gamma$, with well estimated noise levels, allows these realisations to become constrained by the data. 
\par
\begin{figure*}

\begin{tikzpicture}[squarednode/.style={rectangle, rounded corners, draw=black!60, fill=gray!5, very thick, text width=28mm, align=center, node distance=10mm},squarednodesmall/.style={rectangle, rounded corners, draw=black!60, fill=gray!5, very thick, text width=10mm, align=center, node distance=5mm}, emptysmall/.style={rectangle, rounded corners, draw=white!60, fill=gray!5, very thick, text width=5mm, align=center, node distance=2.5mm}]
\node[squarednode]  (start) {Survey shear estimates are converted  into pixels on the sphere, $\gamma^{\textrm{data}}_{12}$ and $\sigma^{\gamma}_{12}$};
\node[squarednode] (1st_tr) [right=of start] {An initial hypothesis of the $E$-mode spherical harmonic amplitudes, $A^{\textrm{hyp}}_{\ell m}$, is made};
\node[squarednode] (1st_hyp) [right=of 1st_tr] {Harmonic inverse transform of hypothesis provides shear field, $\gamma^{\textrm{hyp}}_{12}$};
\node[squarednode] (1st_test) [right=of 1st_hyp] {Calculate likelihood for $\gamma^{\textrm{data}}_{12}$ and $\gamma^{\textrm{hyp}}_{12}$};
\node[squarednode] (1st daugh) [below=of 1st_test] { $A^{\textrm{hyp}}_{\ell m}$ used to produce further hypotheses $A^{\textrm{hyp}, j}_{\ell m}$ };
\node[squarednode] (1st comp) [left=of 1st daugh] {$A^{\textrm{hyp}, j}_{\ell m}$ are transformed to $\gamma^{\textrm{hyp}, j}_{12}$ fields and likelihood calculated};
\node[squarednode] (q) [left=of 1st comp] {Is the fit improved?};
\node[emptysmall] (none) [below=of q] {};
\node[emptysmall] (none2) [right=of none]{};
\node[squarednode] (adopt) [below=of 1st comp] {Best likelihood coefficients adopted as $A^{\textrm{hyp}}_{\ell m}$};
\node[squarednode] (end) [left=of q ] {Output the final $A_{hyp}$};
\node[squarednode] (chi2check) [left=of adopt] {Is $\chi_n^2  < n$?};
\draw [-{Latex[width=3mm, length=2mm]}] (start) to (1st_tr);
\draw [-{Latex[width=3mm, length=2mm]}] (1st_tr) to (1st_hyp);
\draw [-{Latex[width=3mm, length=2mm]}] (1st_hyp) to (1st_test);
\draw [-{Latex[width=3mm, length=2mm]}] (1st_test) to (1st daugh);
\draw [-{Latex[width=3mm, length=2mm]}] (chi2check) to [out=180, in=270]node[anchor=south]{Yes}(end);
\draw [-{Latex[width=3mm, length=2mm]}] (adopt) to [out=0, in =270] (1st daugh);
\draw [-{Latex[width=3mm, length=2mm]}] (chi2check) to node[anchor=north]{No}(adopt);
\draw [-{Latex[width=3mm, length=2mm]}] (q) to node[anchor=south] {No}(end) ;
\draw [-{Latex[width=3mm, length=2mm]}] (q) to [out=270, in=90]node[anchor=east] {Yes}(chi2check) ;
\draw [-{Latex[width=3mm, length=2mm]}] (1st daugh) to (1st comp);
\draw [-{Latex[width=3mm, length=2mm]}] (1st comp) to (q);
\end{tikzpicture}

\caption{Graphical representation of the forward fitting routine, describing the iterative nature of the process. Hypotheses are produced in the harmonic space and compared to the data in real space, gradually increasing the likelihood through subsequent selection of improved hypotheses. B modes are set to zero, and only E modes are hypothesised.}
\label{fig:ff_flow}
\end{figure*}
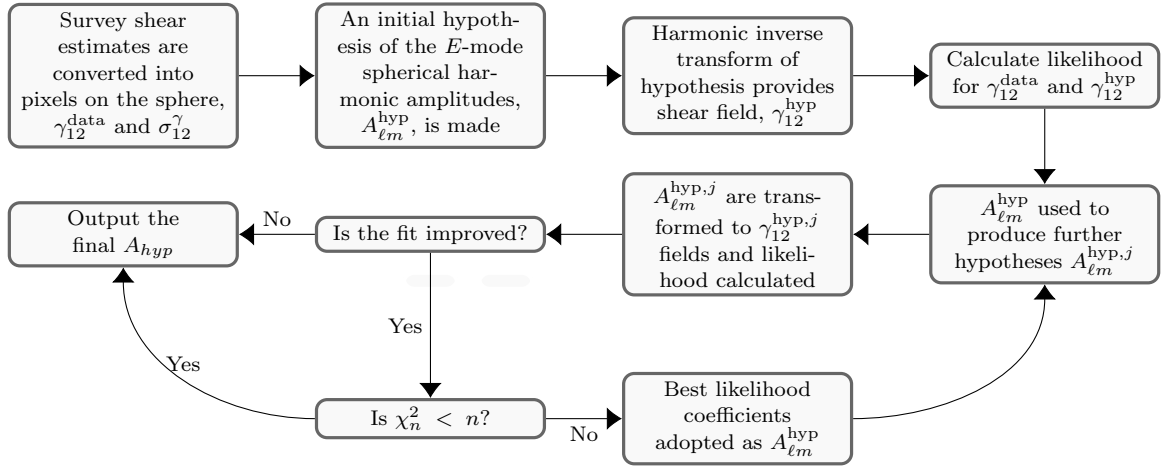

Our technique produces spherical harmonics characterising a full sky of shear observations, which can then be converted back into maps of $\gamma_1$ and $\gamma_2$. An implementation of the method was written in \texttt{python} and made use of the packages \texttt{healpy} and \texttt{numpy}\footnote{https://docs.scipy.org/doc/}. The whole procedure is shown graphically in Figure \ref{fig:ff_flow}, and described below: 
\begin{itemize}
\item An initial shear hypothesis, $A_{\ell m}^{hyp}$, is made for the $E$ modes in harmonic space (up to an $\ell$ of 2 \texttt{nside} -1), and transformed to give its corresponding full sky shear fields $\gamma$ through use of the \texttt{healpy} function \texttt{alm2map}. We choose to make our initial hypothesis through generating an $E$ mode harmonic corresponding to a Gaussian random sky with variance typical of the observed galaxy overdensity field. We do not use the actual overdensity field to inform this initial hypothesis, beyond using it as an estimate of the size of fluctuations in the field. This initial guess is far from the minimum, but early iterations from this guess take large steps down the likelihood surface. 
\item The hypothesised shear field is compared to limited observed data on the sphere $\gamma_{obs}$ with known errors $\sigma_{\gamma}$ by calculating its likelihood. Assuming Gaussian errors and independent measurements, the log-likelihood is proportional to \\
\begin{equation}
\chi^2=\sum_{ i \in \textrm{footprint}_{\textrm{pix}}} \frac{(\gamma_{\textrm{hyp}, i}-\gamma_{\textrm{obs},i)^2}}{\sigma_{\gamma, i}^2} 
,
\label{eq:chi2}
\end{equation}
where footprint$_{pix}$ is the set of pixels contained in the survey footprint. We use this as we find that the diagonal terms of the covariance are dominant, and a full pixel by pixel covariance matrix is prohibitively computationally expensive. When we investigated the covariance of a pixel with its neighbour across simulated maps, we found that the covariance with respect to the immediately neighbouring pixel is at a  value approximately 4.5\% of the variance within the pixel; the corresponding inverse covariance matrix is well approximated as diagonal with elements equal to the reciprocal of the variance. 
We wish to identify a confidence region for convergence pixel values; since equation (\ref{eq:chi2}) follows a $\chi^2_n$ distribution where $n$ is the number of pixels, and we are fitting $n$ parameters (i.e. pixel values), the 68\% confidence region is bounded by $\chi^2/n\simeq 1$.
\item $A_{\ell m}^{hyp}$ serves as the basis for a series of similar $E$ mode harmonics, $A^{j}_{lm}$, produced through randomly perturbing the coefficients of the $m$ modes of a particular randomly chosen $\ell$, by adding a contribution to each $m$ drawn randomly from a Gaussian with standard deviation determined by the power in the current best hypothesis at that $\ell$ mode. We choose to produce 10 perturbed hypotheses, each differing from each other by altering the coefficients at a single $\ell$ mode. 
\item Each of these altered hypotheses $A^{j}_{lm}$ are transformed to produce shear maps $\gamma^{j}$ which are then compared to the data and the corresponding likelihood calculated.
\item The hypothesis with the greatest likelihood is adopted as the new $A_{lm}^{hyp}$ for the next iteration of the routine.
\item The cycle repeats until consistency with the data is found ($\chi^2/n < 1$), or a plateau reached from which no improvements to the fit are made after a threshold number of attempts. For the results in this paper, we chose 500 generations without improvement as the exit condition. In practice no runs required this exit condition.
\end{itemize}

The final $A_{lm}$ can be used directly as $E_{lm}$ in equations (\ref{eq:kap_ell}) and (\ref{eq:alpha_ell}) to produce a convergence map. Repeated runs of the fitting algorithm produce a sampling of consistent maps. We produce a final map from the average of 40 fitted maps; we will show in later tests that this map provides an unbiased reconstruction of the mass distribution, and assess the errors associated with it in Appendix \ref{app:noise_modelling}. \par

We can carry out a full sampling of the likelihood using the Metropolis-Hastings MCMC algorithm \citep{Metropolisdoi:10.1063/1.1699114,  Hastings10.1093/biomet/57.1.97}, which we have tested for one of our \texttt{synfast} all sky simulations, with 10 walkers starting from randomly generated starting points (i.e. shear pixel values from Gaussian random variable with s.d. equal to shear s.d. of data pixels) and each storing a sample every 100 steps. We let these walkers run for $\sim 10^6$ steps and all reached $\chi^2/n \simeq 1$, but few outputted maps were at a $\chi^2$ value much lower than this. As we have a high number of pixels, the volume of possible solutions rapidly decreases as $\chi^2$ is lowered and subsequent steps in the direction of an increased likelihood fit become less and less probable. Due to the 786432 convergence values stored for a full sky map at \texttt{nside}=256, this approach requires Terabytes of storage, which is prohibitive if used for many simulations, as for instance needed in Appendix \ref{app:noise_modelling}. Our approach, averaging only 40 final outcomes of $\chi^2$ descent, allows us to find a point consistent with the mean found from MCMC in an efficient fashion; the mean difference  in a pixel between the MCMC output and our approach is $\simeq 0.24 \sigma$, (where $\sigma$ is the standard deviation of $\kappa$ pixels found in the MCMC run) and there is a Pearson correlation coefficient of 0.92 between the maps made with each method. In addition, we note that via MCMC or direct descent, the maximum likelihood solution takes a prohibitive time to reach; we do not require this solution as our average reconstruction is fully consistent with the true field (see section \ref{sec:gauss_tests}).

The runtime of the routine is predominantly affected by two variables - the size of the perturbations given to the coefficients and the \texttt{nside} resolution of the field being fitted. To optimise the fitting time, we need to minimise the number of steps taken to maximise the likelihood. We find that our routine can perform $\simeq$20,000 generations of 10 hypothesis fields at an \texttt{nside} $=256$ and return an output map in $\simeq$24 hours, using 8 cores on the SCIAMA HPC cluster. The direct inversion is much quicker, taking a maximum of one minute for a field of the same resolution.  Due to the nature of the fitting routine, the direct inversion will always be significantly quicker than the forward fit. \par
Note that other techniques exist using forward fitting, some with more constraining priors. For example, \citet{NJeffrey2018MNRAS.479.2871J} used DES data with Gaussian and sparsity priors to examine how these affected the reconstructions, and found both to improve on the maps made by the direct inversion. \citet{Alsing2017MNRAS.466.3272A} have used Bayesian Hierarchical Models to simultaneously infer cosmological parameters, power spectrum and shear maps for CFHTLenS. This sophisticated approach will be very promising for curved sky maps in the future; our method can be considered to be a step towards this approach for wide areas, focusing on the map-making element only. 

\subsection{Examining the reconstructions}
\label{sec:rec_stat_comp}
\subsubsection{F Statistics}

There are several statistics that we will use to quantify the success of reconstructions, the first of which are:
\begin{equation}
F_{1}=\sqrt[]{\frac{\langle \kappa_{rec}^{2} \rangle}{\langle \kappa_{t}^2 \rangle} } ; \;
F_{2}=\frac{\langle \kappa_t \kappa_{rec} \rangle}{\langle \kappa_t^2 \rangle}
\label{Fstats_eq}
.\end{equation}
$F_1$ measures the consistency with which the amplitude information of the maps is preserved, whereas $F_2$ is sensitive to how well phase information is recovered. A result of unity for both statistics would mean that the reconstruction is managing to perfectly capture both the phase and amplitude information in the map,.  Once the behaviour of our fields' $F_1$ is understood, we may be able to account for any changes in amplitude by applying a corrective multiplicative factor across the map, but if $F_2$ is significantly degraded then this indicates that phase information is lost in the final result. Therefore, we want $F_2$ to be as high quality as possible, and $F_1$ to be well understood.   \par

Furthermore, when we measure these statistics in maps including shape noise, it becomes important to correct for the effect of this noise contribution, which we call ``denoising'' of the statistics. Consider a reconstructed $\kappa$ map, composed of two parts:
\begin{equation}
\kappa_{rec} = \alpha \kappa_{sig} +\kappa_{n}
,
\label{eq:alphakappan}\end{equation}
where $\kappa_{sig}$ is the reconstruction of the true convergence from the true shear, and the $\kappa_{n}$ term encapsulates all other noise effects that alter the convergence from its true value, such as edge effects and noise in the measurement of the shear in pixels. If we can successfully model the noise on the $\kappa$ maps, then we can correct for this noise in $F_1$; our denoised $F_1$ therefore becomes
\begin{equation}
F_{1} = \sqrt[]{F_{1, N}^2 - F_{1, n}^2 }
\label{eq:denoise}
,\end{equation}
where $F_{1, n}$ is the statistic found when producing a map that consists solely of noise, and $F_{1, N}$ is the statistic found from the initial noisy data. We assume that there is no correlation between the convergence $\kappa$ and the noise in that pixel. The denoising procedure serves as an important test of how well we can model the noise across the map, which is of particular importance when we later attempt to reconstruct moments of the true $\kappa$ distribution. A denoising procedure that recovers $F_1$ statistics close to unity  indicates that both the reconstruction and the noise are behaving as expected. 
The $F_2$ statistic does not require any denoising, under the assumption of no correlation between $\kappa_{true}$ and the noise contribution. However, introducing multiplicative systematics into the maps (i.e. $\alpha$ in equation (\ref{eq:alphakappan})) will change the $F_2$ statistic. \par

\subsubsection{Minkowski Functionals}
Another set of statistics of interest are the Minkowski Functionals, which encode the topological information of a map \citep{Mecke:1994ax} and are therefore useful to constrain cosmological models \citep{SchKersBuch_1996dmu..conf..281S,Kers_1996ASPC...94..247K,PetriMF:2015ura} . 
Following the notation of \citet{Hik_2006ApJ...653...11H,DMMF2012MNRAS.419..536M}  the  three Minkowski Functionals ($V_0, V_1, V_2$) for a 2D surface are defined as 
\begin{equation*}
V_0=\int_\Sigma \textrm{d}a,
\end{equation*}
\begin{equation*}
V_1 = \frac{1}{4}\int_{\partial \Sigma} \textrm{d}l,
\end{equation*}
\begin{equation*}
V_2=\frac{1}{2 \pi} \int_{\partial \Sigma} K \textrm{d}l,
\end{equation*}
and represent integrals across the total area, length and curvature characteristic of a given excursion set of the map respectively. We calculate these quantities through the use of angular derivatives on the sky \citep{minksph_doi:10.1046/j.1365-8711.1998.01467.x}, using the formulae listed in Appendix A of \citet{Hik_2006ApJ...653...11H}. Each are normalised by the area of the complete map. We estimate derivatives through finite differences between neighbouring pixels in a similar way to \citet{PetriMF:2015ura} but implemented on the spherical pixel scheme used in these maps. Appendix \ref{app:functionals} gives details of the formulae used to calculate the functionals from a given \texttt{HealPIX} map.\par 

\subsubsection{Pearson Correlation}
Finally, we will also use the Pearson Correlation coefficient, defined as
\begin{equation}
\rho_{X,Y}=\frac{(X-\hat{X})(Y-\hat{Y})}{\sigma_x \sigma_y}
\end{equation}
For a perfect correlation of $X$ and $Y$ fields, this will be exactly unity, and deviations from this value provide a measure of the level to which noise can dominate the signal.This statistic does not need a de-noising approach. In our analysis, we will be using a reconstructed field and a true $\kappa$ field as $X$ and $Y$.\par
\section{Simulations and initial tests}
In this section we will use simulations of observed $\gamma$ fields with known $\kappa$ fields to compare the two reconstruction methods, using the metrics introduced in Section \ref{sec:rec_stat_comp}. In all further maps in this paper, results are for final fitted maps of \texttt{nside}$=256$, with a pixel separation of $~0.22$  deg unless otherwise stated. We use a maximum $\ell$ mode of 2 {\tt nside} - 1 = 511, for sufficient resolution and speed.  Maps are then smoothed with a Gaussian with standard deviation of 20 arcmin. We find that the forward fitting routine consistently performs better than the direct inversion, across a variety of metrics used in this section. \citet{Chang:2017kmv} produced maps to a resolution of \texttt{nside}$=1024$, and results found here are not necessarily directly comparable due to the different resolutions.  
\label{sec:tests}
\subsection{Gaussian map tests}
\label{sec:gauss_tests}
The precision and accuracy of each reconstruction technique needs to be carefully assessed, for both the forward fitting approach and for the direct inversion. There are several different effects that will immediately degrade the reconstruction: the limited survey footprint, and the fact that we observe a shear estimate using a varying number of galaxy ellipticities, which introduces a pixel by pixel variation in the noise properties. \par

To examine these effects, we produced a series of full sky shear and convergence maps using the \texttt{healpy} routine \texttt{synfast}. This routine produces Gaussian random fields from an input power spectrum, which for these tests was that of a flat $\Lambda$CDM Universe characterised by the parameters $\Omega_m = 0.3, \Omega_b = 0.047, h=0.7, \sigma_8 =0.82, w=-1$.  The power spectrum was calculated using \texttt{COSMOSIS} \citep{ZUNTZ201545_COSMOSIS} which utilises the \texttt{CAMB} code \citep{camb-Lewis:2002ah}. A `true' sky convergence distribution is given by the output of \texttt{synfast}, with matching true shears. We fit to 25 different true skies with different noise fields, drawn from the same map of pixel uncertainties matching the Buzzard footprint. For a fairer comparison with our other simulation tests, these uncertainties are calculated for a shear field with an error on $\gamma$ components of $\approx 0.27$ and galaxy number in each pixel obtained from the Buzzard catalogue (mean pixel galaxy count of $\approx 600$) which is described in  section \ref{sec:galaxy_sims}.\par

The forward fitting approach was applied to this simulated data and produced converged fits. Maps were also made by the direct inversion, in order to compare the two approaches. \par  

\begin{figure*}
\includegraphics[width=\textwidth]{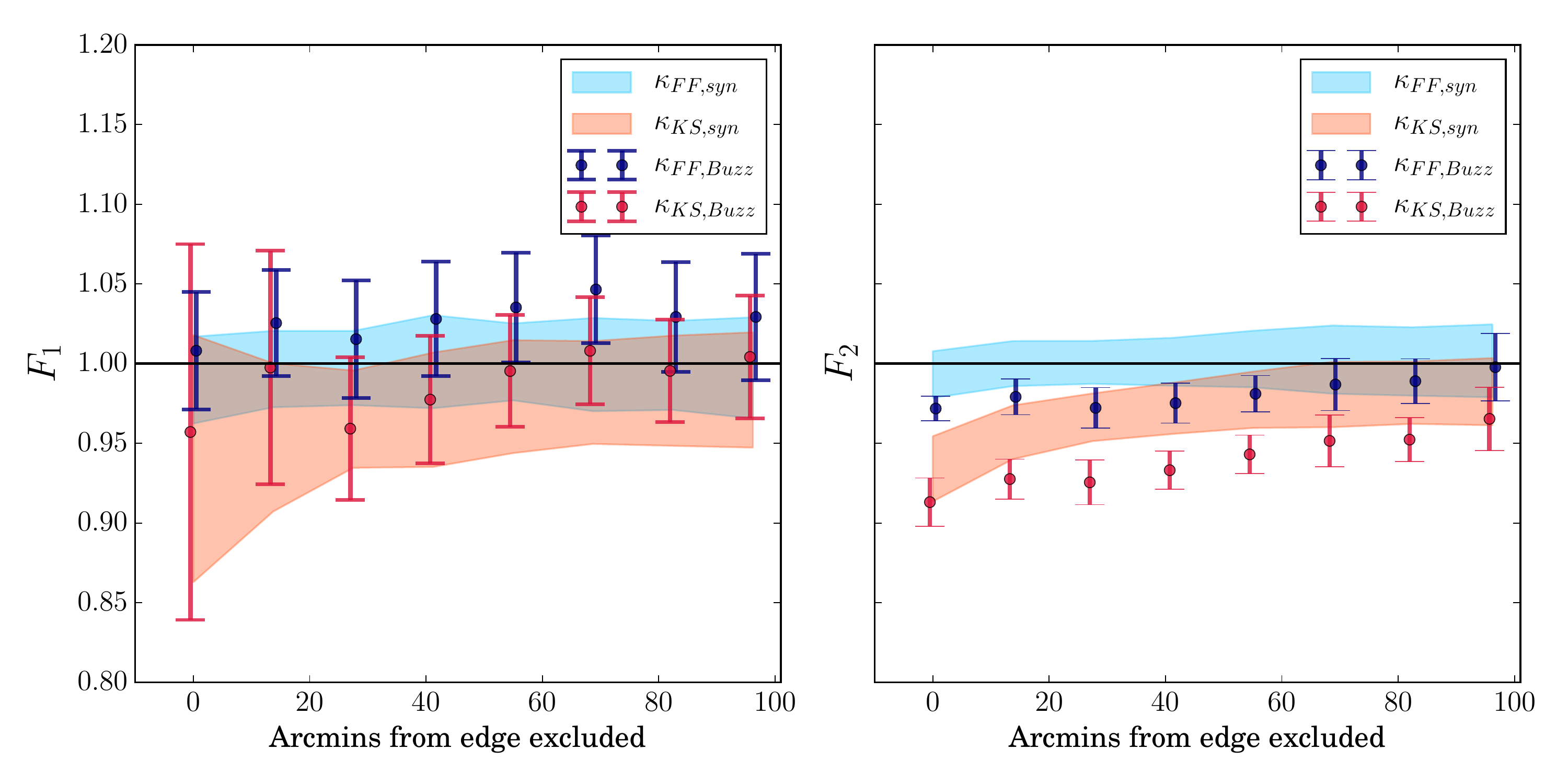}
\caption{The $F$ statistics for the two reconstruction methods in multiple simulation tests. The solid filled regions are for the Gaussian simulations with known noise properties, and the scatter points are for the Buzzard simulations. The horizontal axis represents the width of the pixel range around the edge of the surveyed area which is removed before calculating $F$ for the remaining areas of the footprint. Noise estimates for the Buzzard simulations were found by using the residuals from a fitted sky and a known, Gaussian simulated truth using \texttt{synfast}. The $F_1$ statistic has been de-noised, but the $F_2$ statistic does not require this procedure.}
\label{fig:fstats}
\end{figure*}

\begin{figure}
\includegraphics[width=0.5\textwidth]{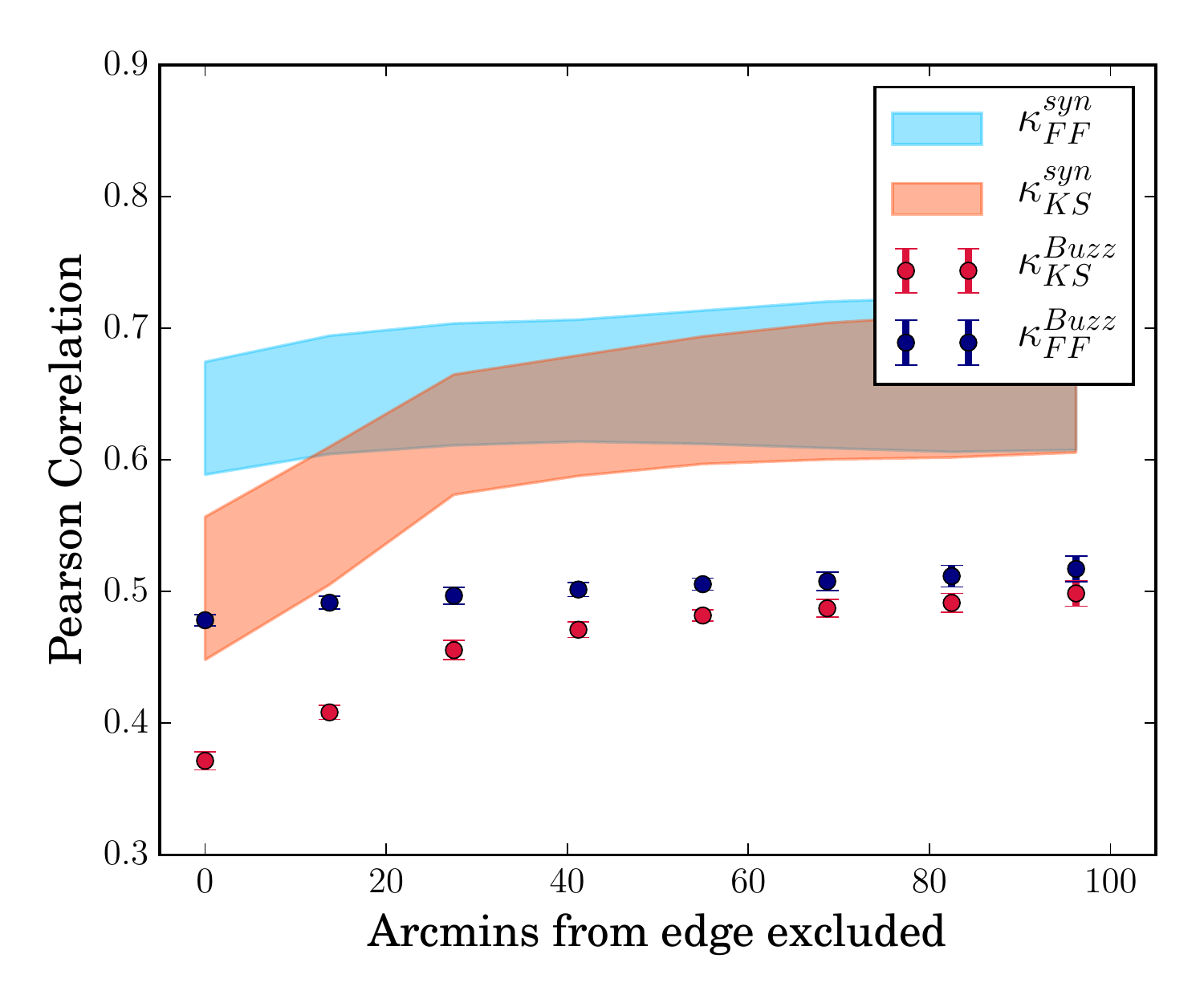}

\caption{The Pearson correlation coefficient between the different reconstruction techniques and the true convergence $\kappa_t$, excluding pixels that fall within a given distance of the edge of the survey. The results for both the simulations using \texttt{synfast} (filled region) and those using the Buzzard simulation are shown. The deterioration of the reconstructions as the edge pixels are included in the measurement of the statistic can be interpreted as the contribution from these more noisy pixels reducing the correlation. Significant improvement in correlation can be seen as the noisier exterior pixels for $\kappa_{KS}$ are excluded, but for $\kappa_{FF}$ the variation due to excluding these pixels is much less pronounced. }
\label{fig:Pearsons}
\end{figure}

\subsubsection{Fidelity metrics}

Figure \ref{fig:fstats} shows how well the $F$ statistics behave for these simulations, with the filled region showing the Gaussian field 1-$\sigma$ spread of results when these are evaluated across the 25 maps. The $F_1$ statistic has been de-noised in this plot and is found to be consistent with 1; this implies that our method of modelling noise is consistent with the true noise within that pixel. For the Kaiser-Squires technique applied to the data, the large errors in these outer pixels cause more scatter in the denoised statistic. \par
Considering the $F_2$ statistic, we see the direct inversion clearly differs from unity, with the reconstruction becoming worse nearer to the edges. By contrast, the forward fitted map consistently preserves the phase information significantly better across the survey area. This can also be seen in the Pearson correlation coefficient in Figure \ref{fig:Pearsons}, where the edge effects mean that the coefficient for the direct inversion is $\approx 0.1$ lower than our method when evaluated over the $22830$ pixels in the footprint. \par

\subsection{Quantifying noise}
The Gaussian maps described in Section \ref{sec:gauss_tests} were also used to estimate the uncertainties on each pixel of the final convergence maps for simulated and real data. Appendix \ref{app:noise_modelling} describes our examination of the noise properties of the final fitted maps. 

In order to estimate the error in any given pixel, we simulate many Gaussian maps and take the difference between the fitted maps and a known truth. Many such difference maps give a sampling of the error distribution in each pixel, with the standard deviation of residuals in a pixel across these mock maps used as the pixel uncertainty estimate. These simulations use the appropriate shear errors in each pixel for the data we are attempting to simulate. We use 25 simulations of noisy skies to produce our errors, and errors for the direct inversion were found using the same maps. This approach was used to calculate errors for all simulation tests and on the Y1 data. In this paper, error refers to the standard deviation of the residuals found in this way, and noise refers to a single map realisation with values drawn from this distribution for each pixel.\par

\subsection{Galaxy survey simulations}
\label{sec:galaxy_sims}
\begin{figure*}
\includegraphics[width=\textwidth]{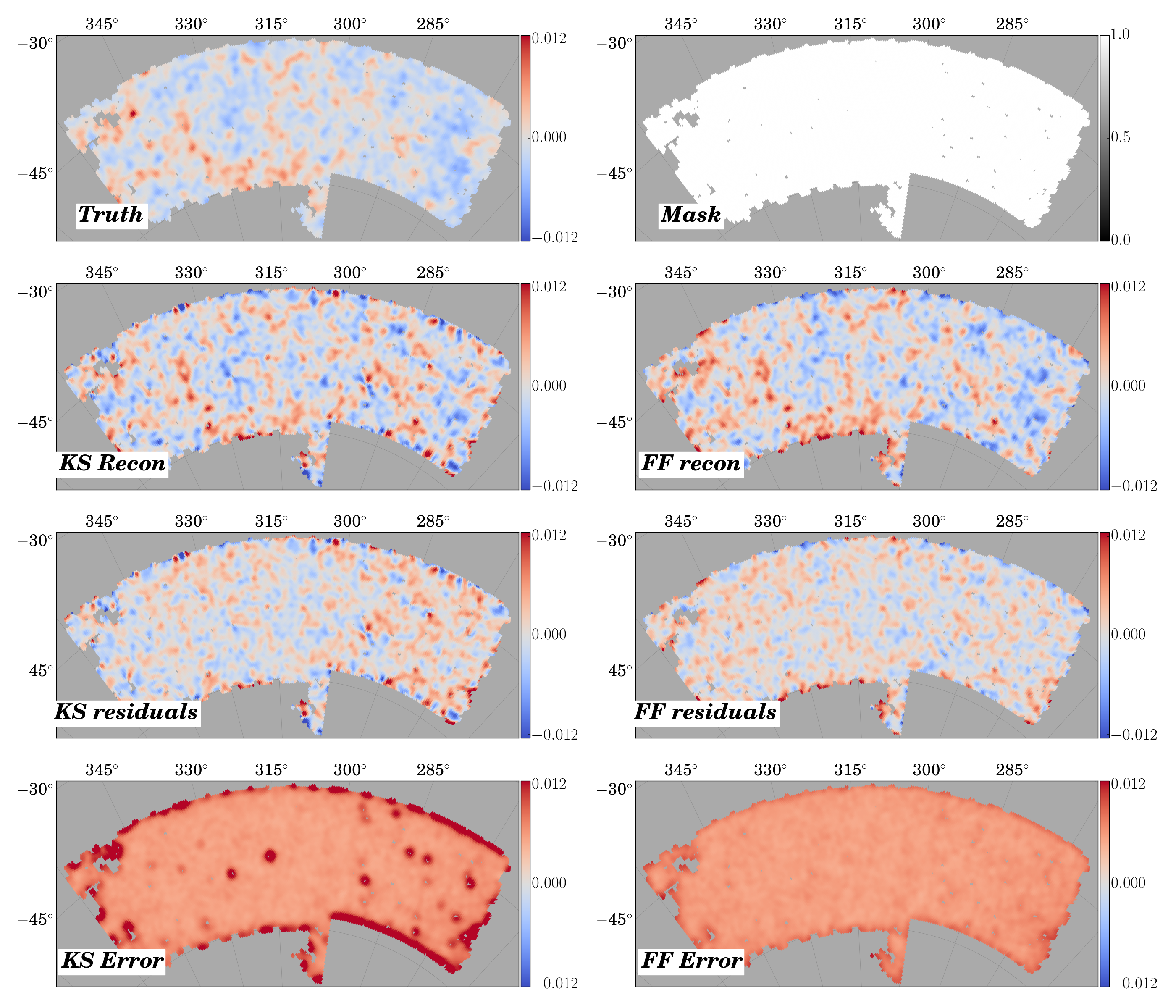}
\caption{Reconstructions of the convergence field for the Buzzard simulation for both the direct inversion (left column) and the fitting method (right column). Smaller errors are present in the edge pixels of the fitted map. The fitting method also has a much more uniform residual map inside the footprint, whereas the direct inversion has a large portion of the survey with slightly larger residuals.  }
\label{fig:BuzzardRecon}
\end{figure*}
Further to the Gaussian simulations, we also test the reconstructions using the Buzzard simulations \citep{DeRose_2019arXiv190102401D}, specifically the ``Buzzard v1.3" mock galaxy catalogues. These catalogues are for 6 simulations resembling the DES Y1 data set, with accompanying galaxy shears, ellipticities and $\kappa$. \par
These were produced via N-body simulations consisting of only dark matter, in a flat $\Lambda CDM$ Universe, through the use of \texttt{LGadget-2} \citep{gadget_doi:10.1111/j.1365-2966.2005.09655.x} with initial conditions from \texttt{2LPTIC} \citep{crocce_doi:10.1111/j.1365-2966.2006.11040.x} and using outputs from CAMB \citep{camb-Lewis:2002ah}. Three boxes sized $1050^3$ , $2600^3$ and $4000^3$ Mpc$^3 h^{-3}$  were simulated using $1400^3$, $2048^3$ and $2048^3$ particles respectively, assuming a background cosmology of $\Omega_m = 0.286$, $\Omega_b = 0.047$, $\sigma_8=0.82$, $h=0.7$, $n_s =0.96$ and $w=-1$. The coarser simulations were used to produce sufficient volume for DES, and the higher resolution output was used to tune smaller scale modelling. It is worth noting that all of these parameters are consistent with the results found in the DES Y1 3$\times$2-pt results \citep{DESY1Cosmo2017arXiv170801530D} even though perfect modelling of cosmological parameters is not necessary to test the mass reconstruction.\par
Using the outputs of these simulations, the empirical \texttt{ADDGALS}  algorithm  (Wechsler+, in prep; de Rose+, in prep) populated the haloes with galaxies, replicating results found with subhalo abundance matching (SHAM) \citep{Conroy_0004-637X-647-1-201,Reddick_2013ApJ...771...30R} by fitting a model to a smaller, higher resolution simulation and applying this model to the larger simulation. \texttt{ADDGALS} simultaneously fits both the distribution of galaxy over-densities and the distribution of $r$-band absolute magnitudes of galaxies, through matching a luminosity function to observed galaxy counts.  
These galaxies are further provided with full SEDs from SDSS DR6 \citep{sdss62008ApJS..175..297A} to produce the $grizY$ magnitudes. \par
Lensing parameters are also computed for the catalogues, in the form of $\gamma$ and $\kappa$ for each galaxy. This is done through the use of the multiple plane ray tracing algorithm Curved-sky-grAvitational Lensing for Cosmological Light conE simulatioNS (CALCLENS; \citep{Becker_doi:10.1093/mnras/stt1352} ). The routine uses projected density fields to produce weak lensing maps, with a resolution of $\simeq6.4$ arcsec. The effects of adding photometric noise, adding shape noise, and imposing cuts similar to those in the data catalogue, as described in section \ref{sec:Dark Energy Survey}, are also accounted for to produce an output similar to that of DESY1. The simulated region is smaller than the full DES Y1 observed region by $\approx 600 $ degrees, and its galaxy density count is lower by $\simeq$20\%. \par

We apply the forward fitting method to the Buzzard data and output a final map, which is an average of many fitted maps consistent with the data. 
We estimate the pixel error distributions using further simulated \texttt{synfast}\citep{healpix_2005ApJ...622..759G} skies with the same footprint as the Buzzard map, and examine the residuals across an ensemble of final fitted maps and the true $\kappa$ map in each case. These simulations serve as a simple model of the lensed Universe that we are observing: a uniform source plane of galaxies which undergo a lensing effect, onto which a shape noise component is added. In reality, the broad redshift bins that we use to produce sufficiently large galaxy counts on the source plane mean that this model is a simplification, but we will see that it provides a sufficient estimate of errors for our purposes.\par

The maps produced for the Buzzard simulation are shown in Figure \ref{fig:BuzzardRecon} and the residuals for the forward fitting method can be seen to be much lower in the outer edge regions, similarly to that found for the Gaussian simulations, although the errors are now slightly larger. This may be due to non-Gaussianities in $\kappa_t$, or due to the extra noise inherent in all of the Buzzard simulated maps, where we do not have an exact value for the true $\kappa$ and shear in a pixel  but instead estimate it from the measured quantities for galaxies within that pixel.

\subsubsection{F statistics}
The F statistics for Buzzard are plotted as the scatter points on Figure \ref{fig:fstats}. In the case of $F_1$, the de-noised statistic can be seen to be consistent with that found for the Gaussian maps, meaning that our estimate of the noise for these Buzzard simulations is reliable. For $F_2$ both techniques appear to behave slightly worse than in the previous simulations, due to additional noise in Buzzard. The slightly lower $F_2$ and slightly high $F_1$ indicate that this map has larger residuals than that typically found for the Gaussian maps, but these are still consistent with our expectations.\par

\subsubsection{Pearson correlation}
The Pearson correlation coefficient measurements can be seen in Figure \ref{fig:Pearsons} for both reconstructions and for both methods of simulating the data. The \texttt{synfast} simulations have a higher coefficient for both techniques, and each simulation shows a similar shape when comparing the same reconstruction methods. It is apparent that the statistics found for the fitted maps are significantly less affected by removing outer pixels than for the direct inversion, suggesting the presence of excess noise in the edge pixels of the direct inversion. \par

\subsection{Moments}
\label{sec:Moments}
\begin{figure}
\includegraphics[width=0.5\textwidth]{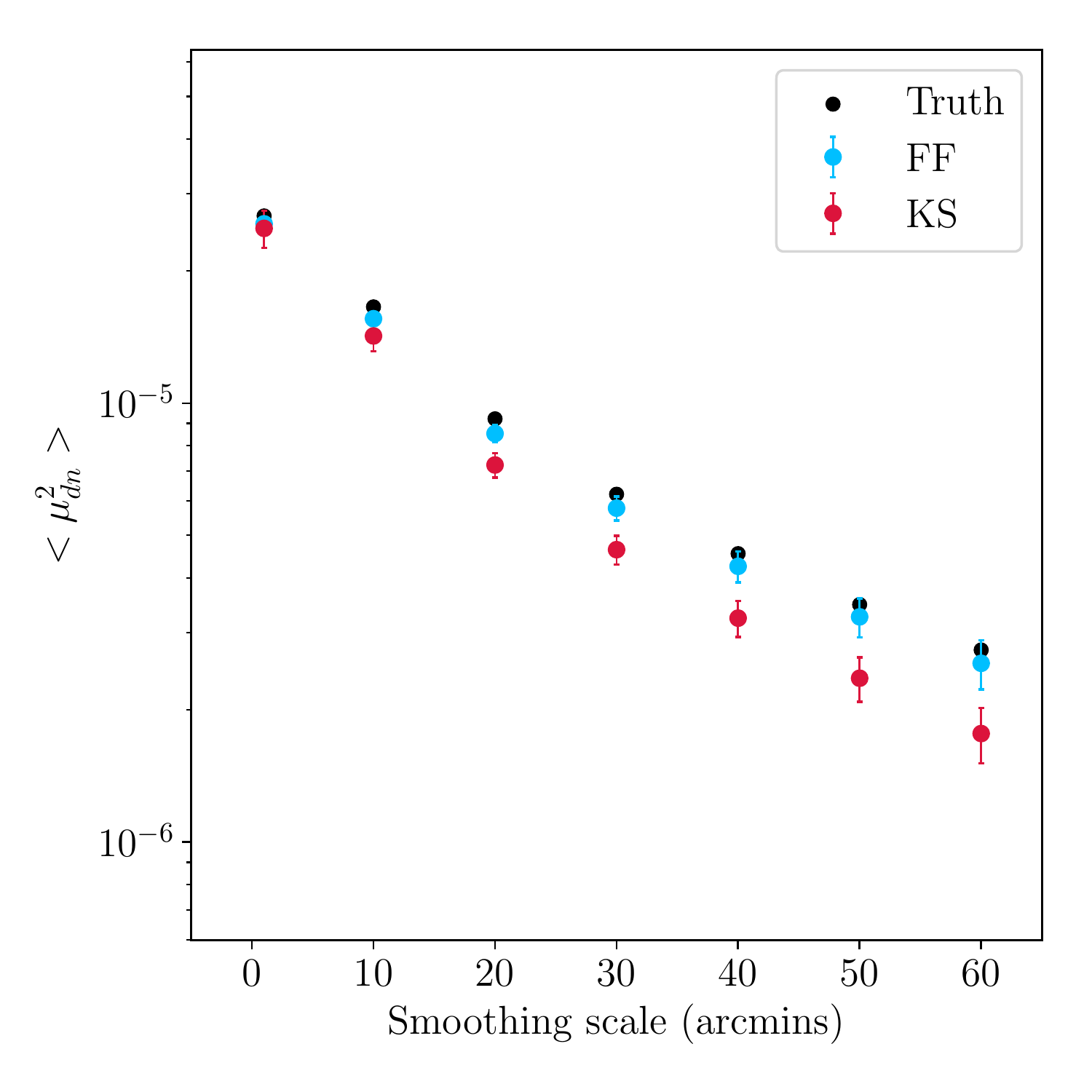}
\caption{ Reconstruction of the second moment of the $\kappa$ field from Equation \ref{eq:denoising_law} for the forward fitting and direct inversion methods, together with true values, for a Buzzard simulation.}
\label{fig:moments}
\end{figure}
Summary statistics such as moments can be very useful probes to test the accuracy of our final maps. Beyond characterising the distribution, the moments as a function of the smoothing scale have also been shown to be useful for constraining the underlying cosmology.  \par
When measuring these moments, we are measuring a combination of the true convergence moment and a noise term. We correct for the noise following the technique used in \citet{VanWae_doi:10.1093/mnras/stt971} and the methodology introduced in equation \ref{eq:denoise}, such that for the second moment the de-noising procedure is:
\begin{equation}
\langle (\mu_{dn})^{2}\rangle = \langle (\mu_{N})^2 \rangle - \langle(\mu_{n})^2 \rangle 
\label{eq:denoising_law}
,\end{equation}
where $\mu_{n}$ is the $\kappa$ field found for the noise exclusively, $\mu_{dn}$ is the denoised $\kappa$,  and $\mu_N$ is the moment found from the noisy field. 

Figure \ref{fig:moments} shows the reconstruction of these de-noised moments for both methodologies on one of the Buzzard simulations. At large smoothing scales, the edge effects from the direct inversion become more and more significant such that the moments become increasingly biased away from the truth. The forward fitting approach does not have this level of significant error localised around the edge of the map, so it can reconstruct the moment more reliably to higher smoothing scales. At higher pixel resolution, these edge effects will become less significant as edge pixels account for a smaller fraction of the total pixel count.

\subsection{PDF}
\label{sec:PDFs}
\begin{figure*}
\includegraphics[width=\textwidth]{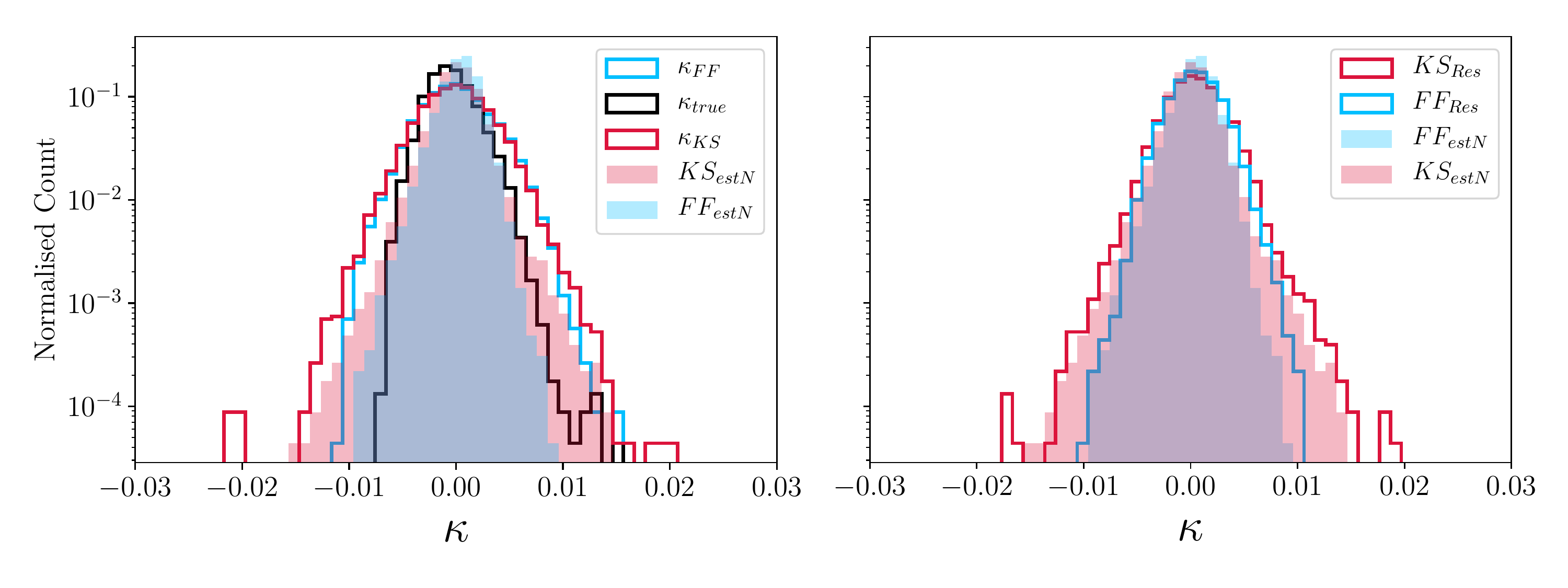}
\caption{PDFs of the reconstruction techniques' convergence fields compared to the true distribution for the smoothed Buzzard simulation shown in Figure \ref{fig:BuzzardRecon}. The left panel shows the distribution of the $\kappa$ in pixels (solid lines), together with the estimated noise distribution (filled region). The solid black line is the true PDF that both reconstructions are attempting to recover. The right panel examines the reliability of the estimates used for the noise distribution, with solid fill areas denoting the true residuals and the solid lines denoting the noise estimates used in the left panel. }
\label{fig:pdfs}
\end{figure*}
The PDFs of the convergence distributions serve as a further probe of the reconstruction, and these are shown in Figure \ref{fig:pdfs}. The large contribution of the noise added to the data can be seen to cause the final PDF to differ significantly from the underlying, non-Gaussian true $\kappa$ PDF. The PDF of our estimated noise distribution is also shown, as well as the PDF of the exact differences between the reconstructions and the truth. These estimates of the noise and the true residuals can be seen to be in good agreement, indicating a good understanding of the noise properties of both methods. Further, the PDF of the $\kappa$ distribution in the forward fitting method manages to retain more of the non-Gaussian shape and does not find an excess of high $\kappa$ values in edge pixels. The forward fitting method finds fewer of the large, negative convergence values which are induced by the noise but reconstructs a similar number of higher positive $\kappa$ peaks to the direct inversion. Some over-dense regions will be real density contrasts, for which both techniques should agree, and some will be the large edge effects that are only present in the direct inversion. \par
We also examine the agreement of these two distributions using the Jensen-Shannon divergence $D_{JS}$, a quantity designed for such a comparison in an information theory context. This is defined, for two distributions A(x) and B(x) as 
\begin{equation}
D_{JS}(A,B)=\frac{1}{2} (D_{KL}(A||M) + D_{KL}(B||M))
,\end{equation}
where 
\begin{equation}
M=\frac{1}{2}(A+B)
,\end{equation}
and $D_{KL}$ is the Kullback Leibler divergence, defined for discrete bins as 
\begin{equation}
D_{KL}(A||B)=\sum_x A(x) \log \frac{A(x)}{B(x)}
\end{equation}
We work in logarithms of base $2$, which means that the Jensen-Shannon divergence can be in the range $0 - 1$, with $0$ indicating identical distributions and $1$ meaning completely different distributions. This provides a way to quantify the similarity between the PDFs of each reconstruction and the known, true $\kappa$ PDF. In the unsmoothed $\kappa$ fields, we find that for the direct inversion  $D_{JS} = 0.349$, whereas for the forward fitted map $D_{JS}=0.308$. In the smoothed maps shown in Figure \ref{fig:BuzzardRecon}, the result for the direct inversion is $D_{JS} = 0.115$, compared to $D_{JS}= 0.103$ for the forward fitted maps. These values show that while both methods are finding rather similar distributions to the desired, true PDF, the forward fitted maps are performing slightly better. We can interpret this result as showing that the forward fitted map has a reduced noise component. From inspection of the PDFs, it can be seen that this is indeed the case in the tails of the distribution.

\subsection{Minkowski functionals}
\label{sec:functional_sims}
\begin{figure*}
\centering
\includegraphics[width=\textwidth]{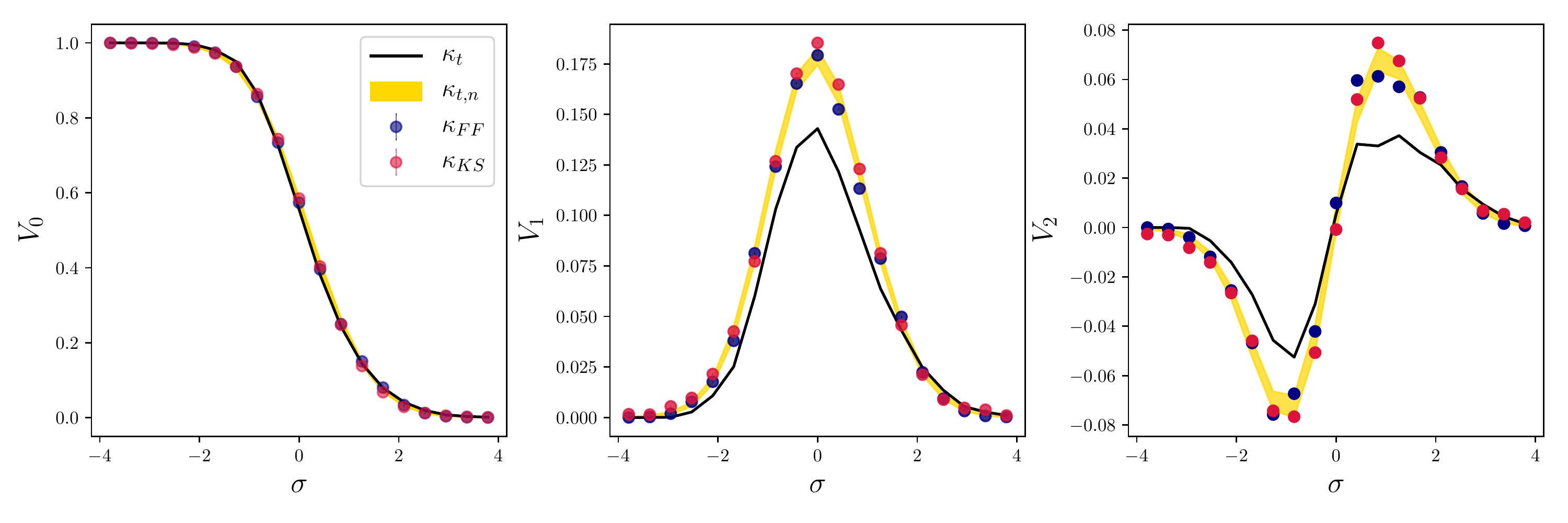}
\caption{The Minkowski functionals for a Buzzard simulation used to test the $\kappa$ map reconstructions, as a function of excursion. In this case, $\sigma$  indicates the number of standard deviations of the final $\kappa$ map and not the error in a pixel. The black line shows the functionals for the true $\kappa$ field and the yellow field region shows the 68\% confidence region for functionals expected when a simple Gaussian noise field is added to this. The Minkowski functionals measure the topology of the maps and perfect agreement between truth and reconstruction across all 3 statistics would mean that all topological information is being retained by the reconstruction. }
\label{fig:functional_buzztest}
\end{figure*}

Figure \ref{fig:functional_buzztest} shows the Minkowski functionals reconstructed for the Buzzard simulation. The solid black line shows the functionals for the $\kappa_t$ field, and the yellow region shows those that we would expect to construct for this field in the presence of noise. The deviations between the functionals for $\kappa_t$ and those in the presence of noise are greatest within a few standard deviations from the mean; this standard deviation approximately coincides with the noise level of these maps. This is expected, as the noise adds artificial peaks of this amplitude which will change the functionals; reducing the noise level would make it easier to distinguish between cosmologies.  Particularly in the direct inversion, the edge effects will introduce some large $\kappa$ values into the distribution and this will alter the standard deviation of the map and consequentially bias the functional measurement. This means that when we attempt to constrain cosmology with these measurements, the reconstruction technique used needs to be taken into account and adequately modelled. \par

We now assess the significance of the agreement between the functionals measured for the reconstructions, and the expectation from the true field with added Gaussian noise. We calculate
\begin{equation}
\chi^2 = (d-t)\textrm{Cov}^{-1}(d-t)^T 
,\end{equation}
where $d$ is the vector of functionals calculated for a reconstruction, $t$ is the vector of functionals for the true $\kappa$ field and Cov is the covariance matrix calculated by examining the covariance between these functional bins when noise is added to the true $\kappa$. This noise is created by generating maps of ellipticity noise, i.e for a pixel $i$, noise is found by randomly sampling from a Gaussian with width according to the error on shear in that pixel. For empty regions of the sky, we fill these areas with noise with the average shear standard deviation of the observed region. These shear maps are then used to produce full sky $\kappa$ maps through the direct inversion. Edge effects are absent from these maps, as the $\gamma$ noise fields are on the full sky, and as such provide a best case scenario for the noise on our final $\kappa$ maps. We are only considering shape noise in our covariance, although we anticipate other contributions to be sub-dominant. Initial tests using Gaussian simulated fields supported this assumption. \par
For binning functionals into $n$ bins, we produce $4n$ of these noise maps. Each noise is then added to the true $\kappa_t$ field and the functionals for each bin are calculated.  We use this ensemble of functionals to measure covariances between bins for our $\chi^2$ calculation, through repeatedly producing new noisy versions of $\kappa_t$ fields.  \par
This approach gives us a way to compare both techniques' abilities to reconstruct the functionals, taking into account the level of shape noise in the data. For the Buzzard simulation results shown in Figure \ref{fig:functional_buzztest}, the reduced $\chi^2$ results for $(v_0, v_1, v_2)$ give $(75/19, 120/19, 73/19)$ for the direct inversion and $(48/19, 38/19, 38/19)$ for the forward fitting routine. This significant improvement shows the impact that edge effects can have on final results for direct inversion; accounting for this is important, as it can bias the final result. \par

\subsection{Comparing residuals}
\label{sec:residuals_sims}
\begin{figure*}
\centering
\includegraphics[width=\textwidth]{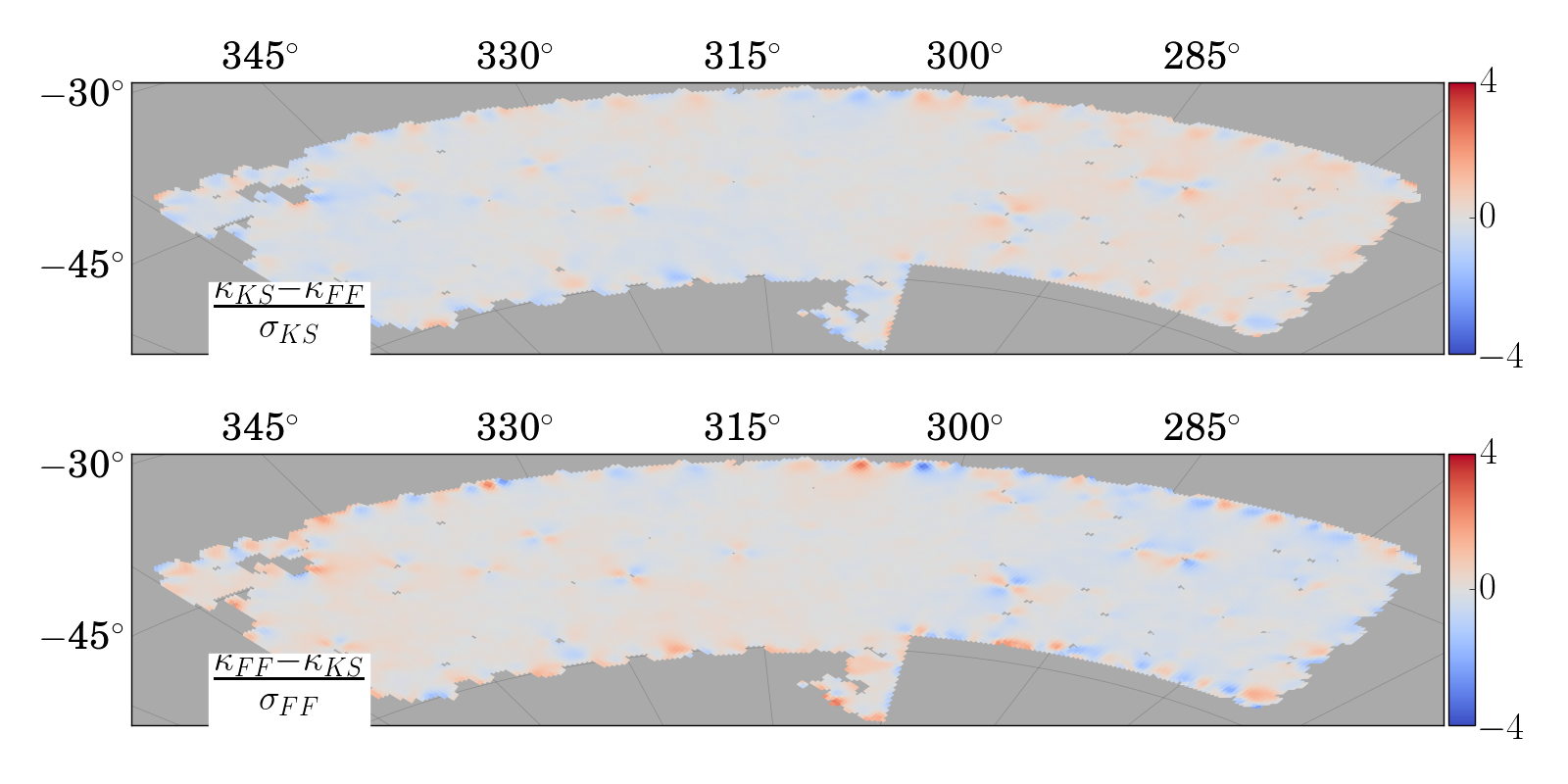}
\caption{The difference between $\kappa$ maps found for both reconstruction techniques applied to one of the Buzzard simulations. As each map technique has different errors in each pixel, the level of tension in terms of standard deviations in a pixel depends upon which map is used as a reference. Disagreements between the direct inversion and the forward fit around the survey edge are of larger significance in terms of $\sigma_{FF}$, indicating that the forward fit is highlighting these particular pixels in the Kaiser-Squires map as being due to edge effects and not caused by the shear signal in the observed region. }
\label{fig:BuzzRecComp}
\end{figure*}
The two reconstruction techniques produce different $\kappa$ maps for the same data, so an interesting comparison is to examine the regions in which they disagree and see what can be learned in these areas. We examine the differences between two maps in ratio to the error in each map in Figure \ref{fig:BuzzRecComp}. Firstly, by visual inspection, the $\kappa_{KS}$ map can be seen to be in much more significant disagreement with the $\kappa_{FF}$ map when differences are taken in terms of $\sigma_{FF}$ than $\sigma_{KS}$. This inconsistency is another illustration of the noise introduced by the Kaiser Squires technique when directly applied to the data. Further to this, considering the signal to noise of these differences in each pixel can highlight tensions between the maps. The two methods are in agreement across large areas of the centre of the footprint, but the edge effects present in the direct inversion  $\kappa_{KS}$ map mean that there is a tension with the forward fitted map. In the 750 pixels where there is a disagreement of $ 2 \sigma_{FF}$ or more, the forwards fitting method has a smaller residual with $\kappa_t$ than the direct inversion in $ 81 \%$ of pixels.

\section{DES Y1 results}
\label{sec:applying_routine}
Following the successful reconstructions of the convergence maps on simulations, we apply the technique to real data from the Dark Energy Survey \citep{DES_summary_doi:10.1142/S0217751X05025917}.

\subsection{Data}
\label{sec:Dark Energy Survey}
The Dark Energy Survey is a photometric survey using the Dark Energy Camera \citep{DECam_1538-3881-150-5-150} on the Blanco telescope, a 4m telescope at the Cerro Tololo Inter-American Observatory in Chile. Observations are taken in five bands (\textit{grizY}).
This work makes use of the data from the first full year of observations, also known as the DES Y1 cosmology data set, or Y1A1 GOLD \citep{Drlica-Wagner:2017tkk}. This footprint is the one best suited to the application of the mass mapping techniques to date, as it spans a substantial survey area of $~1800$ deg$^2$ and contains sources up to a redshift of $\simeq1.3$. 
\subsection{Weak lensing catalogues}

We use the \texttt{Metacalibration} catalogue, which is described in \citet{MCAL1_2017arXiv170202600H} and \citet{MCAL_22017ApJ...841...24S}. This method for calibrating lensing measurements uses the data itself, as opposed to simulations of galaxies, by applying a known shear to deconvolved galaxy images. These galaxy shapes are measured using a Gaussian profile, fit with \texttt{ngmix} \citep{ngmix_2015ascl.soft08008S} and a comparison between the measurement of the change in galaxy shape and the known applied shear gives the \textit{response}. There are three biases associated with this technique: a multiplicative bias $m$, an additive bias $\alpha$ from the PSF ellipticity, and an additive bias $\beta$ from the error on the PSF ellipticity. \citet{Zuntz:2017pso} found these to be $m = 1.2 \pm 1.3\% $, $\alpha \approx 0$ and $\beta \approx -1$. \texttt{Metacalibration} in DES uses images from $r, i$ and $z$ bands and the code is available publicly with the \texttt{ngmix} routines. \footnote{https://github.com/esheldon/ngmix}     \par 
The Dark Energy Survey also used the \texttt{im3shape} routine for producing the weak lensing catalogues and both catalogues were examined in \citet{Zuntz:2017pso}. The $\kappa$ maps produced from both catalogues were  compared in \citet{Chang:2017kmv} and found to be consistent with each other. We choose to produce maps solely on the \texttt{MetaCalibration} catalogue, as this has reduced systematics and a larger galaxy count; there are $\approx$ 34,800,000 galaxies in the \texttt{Metacalibration} catalogue, compared to $\approx$ 21,900,000 in \texttt{Im3shape}, due to the latter only fitting objects in the $r$ band.

\subsection{Redshift binning}
Redshift information in the DES Y1 data catalogue is found by implementing a Bayesian Photometric Redshift (BPZ) algorithm as used in \citet{Benitez_2000ApJ...536..571B} and \citet{Coe_2006AJ....132..926C}, making use of galaxy templates from  \citet{BruCha_doi:10.1046/j.1365-8711.2003.06897.x, Kinney_1996ApJ...467...38K, CoWu_1980ApJS...43..393C} to produce a posterior distribution of the redshift of the observed galaxy. We use the mean of this distribution in selecting the galaxies in a redshift range of 0.3 to 1.2, which are chosen as our source galaxies. We use the redshifts calculated using the \texttt{MetaCalibration} photometry. The full routine is described in \citet{Hoyle:2017mee}, and the use of cross correlation redshifts was tested in \citet{Gatti:2017hmb, Davis:2017dlg}.  \texttt{RedMAGiC} \citep{Redmagic10.1093/mnras/stw1281} galaxies were used as a reference sample for the redshifts, which were in turn calibrated through comparison with BOSS galaxies \citep{Cawthon:2017qxi}. 

\subsection{The DES Y1 fitted maps}
\begin{figure*}
\includegraphics[width=\textwidth]{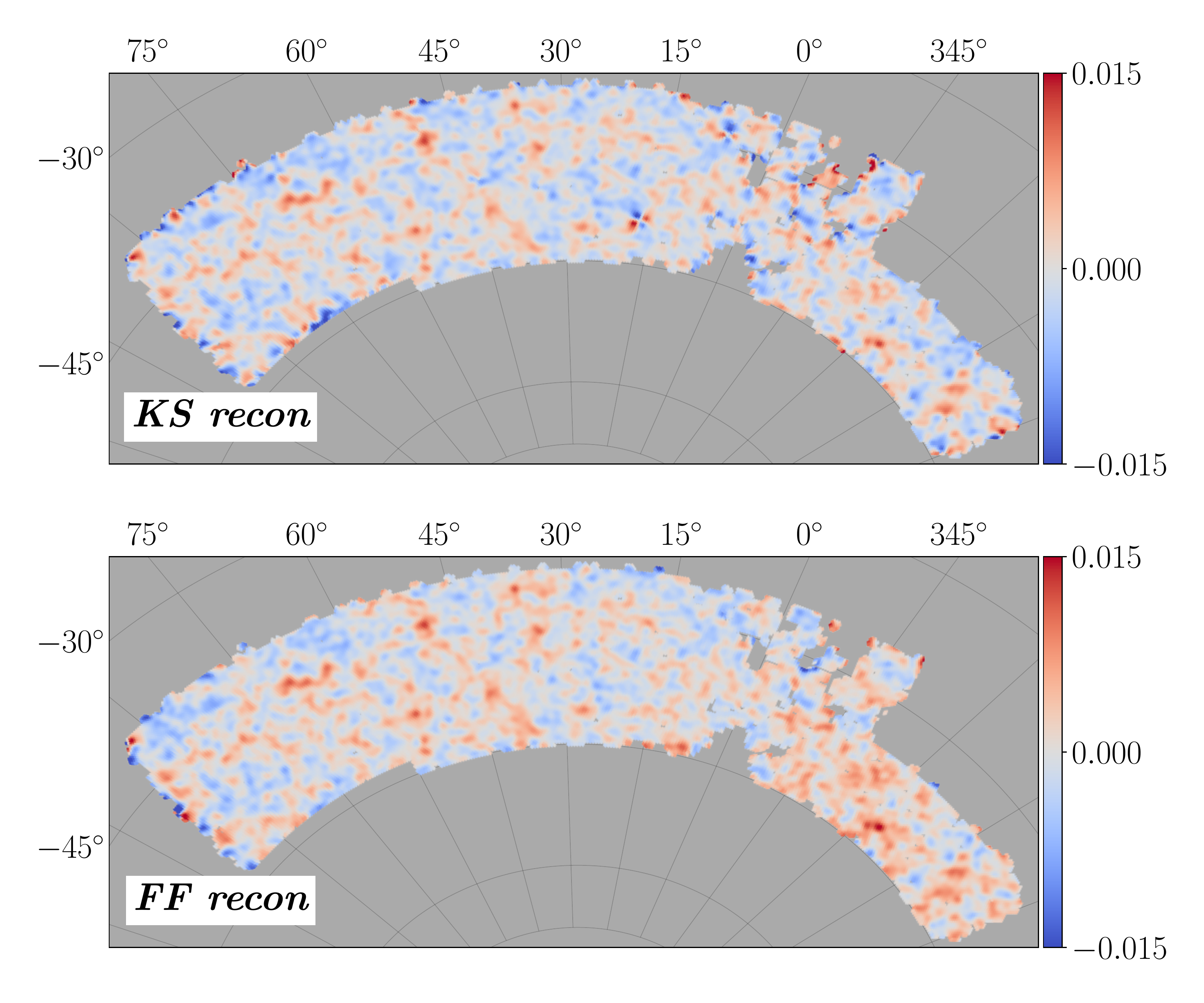}
\caption{The final fitted maps for the DES Y1 data for both 
the direct inversion and the forward fitting method, using 
\texttt{nside}$=256$. The area of complicated masking at 
declination $ < 5\deg$  was not used in our previous wide field mass map paper 
due to the 
footprint introducing significant edge effects.}
\label{fig:y1datamaps}
\end{figure*}

The  fitted maps for the DES Y1 data are shown in Figure \ref{fig:y1datamaps} for the forwards fitting and the Kaiser Squires methods, applied to the full area of the \texttt{MetaCalibration} catalogue. The maps presented here are for the redshift range $0.2 < z <1.3$, which gives the least noisy shear measurements \cite{Chang:2017kmv}. These maps are visually very similar in the inner regions of the map, where peaks in one map frequently coincide with peaks in the other map.  The edge effects are apparent in the reconstruction of the direct inversion in the area of complicated masking (below $5 \deg$ in right ascension) whereas such high $\kappa$ pixels are not produced in the fitted version of the map. This behaviour is similar to that seen in the previous simulations in section \ref{sec:tests}. \par

\begin{figure*}
\centering

\includegraphics[width=\textwidth]{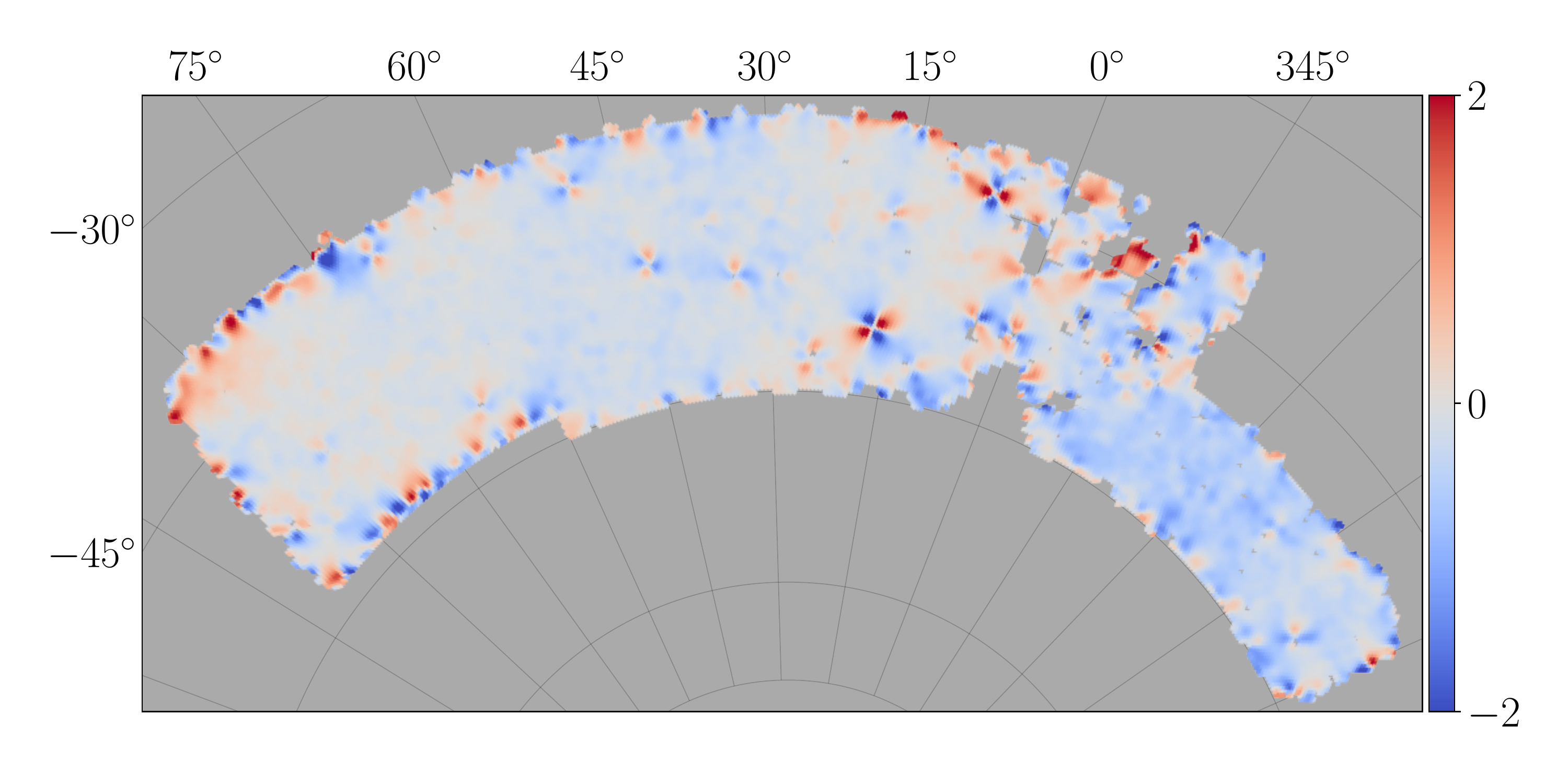}

\caption{The residuals between the two reconstruction techniques, scaled in terms of the error on the forward fitting method in each pixel, and smoothed with a Gaussian of $\sigma=20$ arcmin. As expected, the most significant differences arise in the areas around the edges and in areas of complicated masking.}
\label{fig:KSFF_diffs}
\end{figure*}
\begin{figure*}
\includegraphics[width=\textwidth]{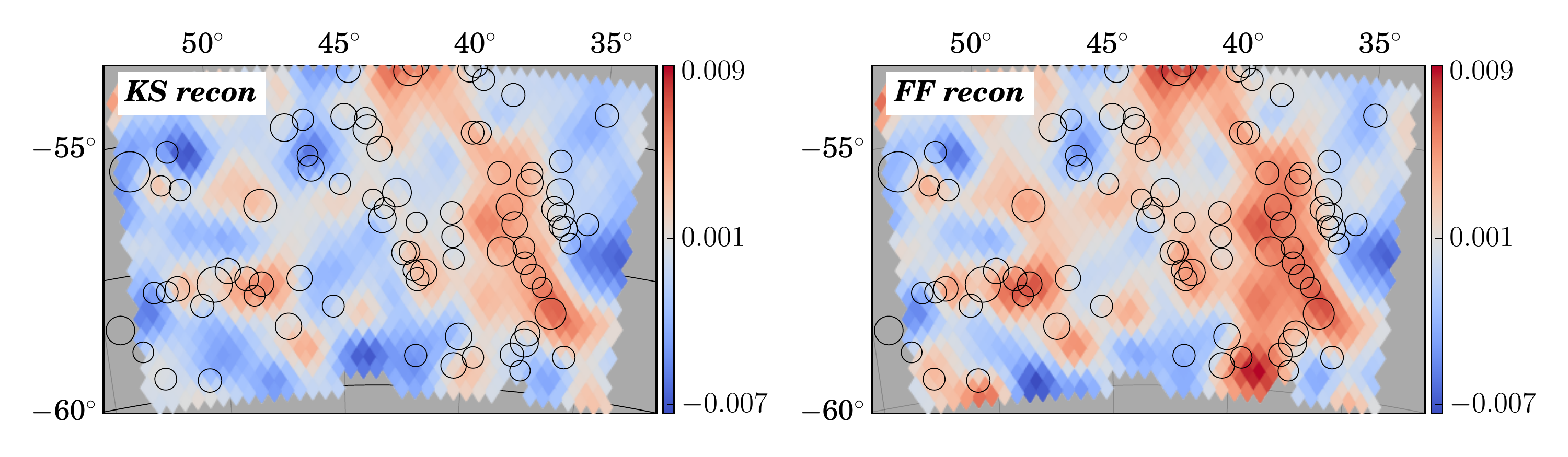}
\caption{A section of the reconstructed maps, with \texttt{RedMaPPer} clusters plotted in black circles, with a radius scaling as the cluster richness $\lambda$, which is an indicator of cluster mass. Clusters were selected in the redshift range $ 0.1 < z < 0.4$. }
\label{fig:clusters}
\end{figure*}

Figure \ref{fig:KSFF_diffs} shows residuals between the reconstructions. 
The direct inversion performs well for $\kappa$ reconstructions well inside the survey footprint, and we see that both methods are working in a similar way in this area as they were in the simulations. In the outer pixels, the differences are typical of those that we would expect from the edge effects, where large $\kappa$ noises and biases are introduced.  The smaller errors of the fitting routine in these edge pixels means that these noisy pixels in the direct inversion are in tension with the fitted results.  In more complicated regions of the mask, there are more frequent disagreements between the two methods, and our tests in section \ref{sec:tests} imply that the forwards fitting method is more suited to these areas. \par
Figure \ref{fig:clusters} shows an area of both reconstructed fields with \texttt{RedMaPPer} clusters \citep{Redmapper_0067-0049-224-1-1} plotted on top. This region is on the southern edge of the survey area. Both maps produce fields of similar morphology, but the forward fitted map has higher peaks due to lacking the edge biases introduced by the direct inversion. Some clusters follow the structure picked out by the maps, with areas of over-density being more frequently populated by clusters. \par 

\subsection{Systematics}

\begin{figure}
\includegraphics[width=0.5\textwidth]{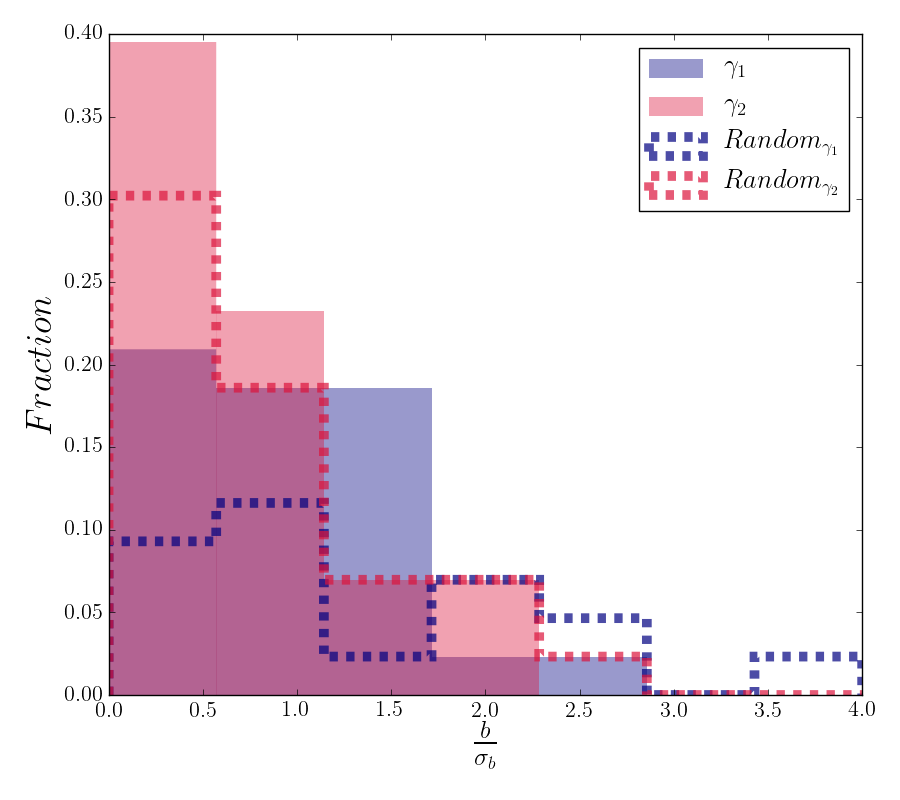}
\caption{The distribution of the magnitude of systematics correlations found for both the Y1 data and for random maps generated using the Buzzard simulations. The latter give an indication of the level of correlation that can arise from random, uncorrelated maps, and therefore give an estimate of an acceptable level of correlation to be found with the Y1 maps.  }
\label{fig:y1systexp_b}
\end{figure}

It is important to correlate these maps with other quantities that are expected to be uncorrelated with the shear in a pixel, to test whether our measurements are free from systematic effects. A significant correlation between our maps and another quantity would indicate how we are not solely inferring mass associated with weak lensing shear, but have a signal contaminated by another source. \par
We will examine the correlation between the observed shear measurements and other observed quantities which could plausibly cause systematic effects, such as the PSF, the airmass, the sky noise and background, across the \textit{griz} filters. Using maps $M^s$ of these observed parameters \citep{systematics2016ApJS..226...24L}, we follow the approach used in \citet{Chang:2017kmv}. Each pixel in the map is assigned a value $n$ between 1 and 10, determined by which decile in the range for that systematic that it belongs to. We then take the average shear value of all pixels of a given $n$, $\hat\gamma_n$. A first degree polynomial is fitted between $n$ and $\hat\gamma_n$, with y intercept $a$ and gradient $b$. The gradient found in this way will therefore indicate a correlation between the shear estimate and the systematic, and a lack of correlation will be indicated by a gradient consistent with 0. Errors on the gradient are found by jackknife resampling. \par
Figure \ref{fig:y1systexp_b} shows the extent of these correlations found for the Y1 data; we see that the correlation between the Y1 data and systematics is no stronger than that found between the systematics maps and a completely uncorrelated map (the Buzzard map).

\subsection{DES Y1 Minkowski functionals}
\label{sec:mink_measurements}
\begin{figure*}
\includegraphics[width=\textwidth]{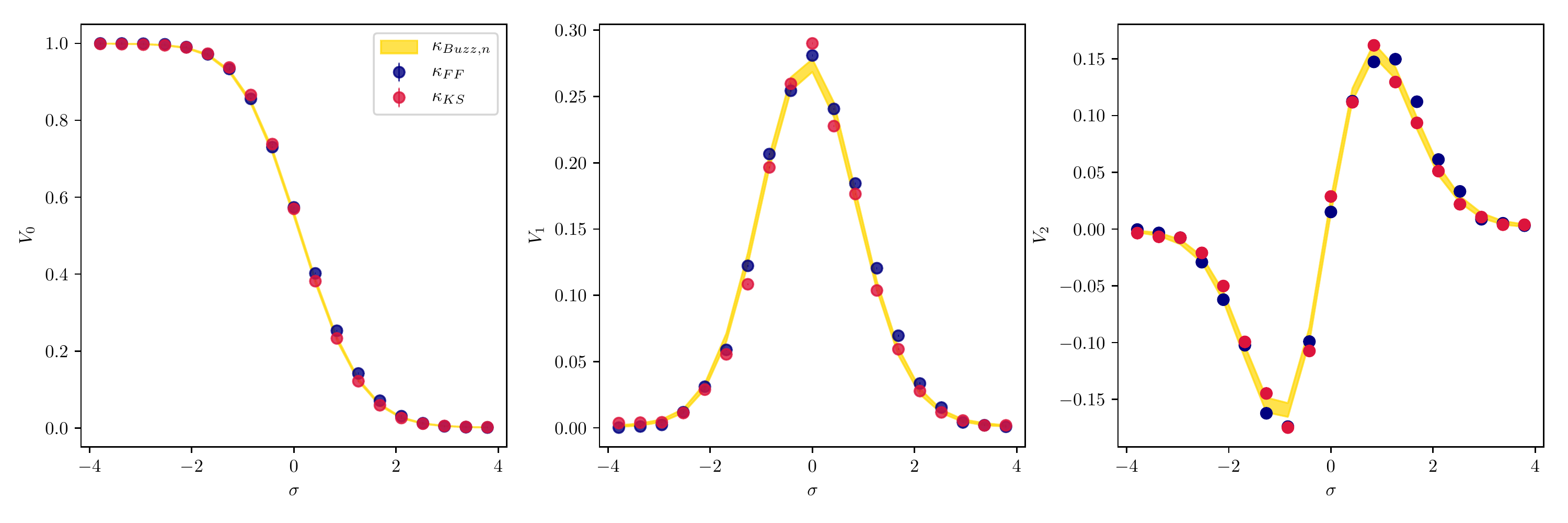}
\caption{The Minkowski functionals for DES Y1 mass maps, found through both the direct inversion and the forward fitting approach.  The solid yellow region shows the expected functionals for a Buzzard field with DESY1-like noise. }
\label{fig:Y1functionals}
\end{figure*}

Following the examination of whether the two reconstruction methods can produce reliable Minkowski functionals in Section \ref{sec:tests}, we can also examine the functionals that describe our final Y1 mass maps. We measure $V_0, V_1$ and $V_2$ (using the methodology in Appendix \ref{app:functionals}) for the final Y1 convergence maps shown in Figure \ref{fig:y1datamaps}. We present these measurements in Figure 
\ref{fig:Y1functionals}, also showing the measurements and uncertainties found for these quantities in the Buzzard simulation. 

As we are only comparing one particular realisation of the Buzzard simulations with the Y1 results, we cannot infer cosmological information from this plot, as we don't know how representative the (finite footprint) Buzzard measurements are of the functionals expected for the Buzzard cosmology, or indeed of the real Universe. However, we see that observed Y1 Minkowski functionals appear similar to those in Buzzard. We will engage in a cosmological analysis of these results in a further paper, where noise will require careful treatment depending on the method used (i.e. direct inversion or fitting), as shown in our earlier analysis in Section \ref{sec:functional_sims}. 

\vspace{0.9cm}

\section{Conclusion}
\label{sec:conclusion}
Weak lensing is one of the leading probes in late time cosmology. Using lensing measurements across large fractions of the sky, it is possible to map the matter distribution on increasingly large scales. We have presented an alternative approach to using the Kaiser Squires formalism to produce wide field mass maps with weak lensing, through producing hypothesis all-sky shear fields which we fit to the data, aggregating over compatible maps. \par
We have tested the forwards fitting approach with simulations, and have found that in each case the resulting maps have reduced residuals compared to the direct inversion method, as well as exhibiting better agreement for the 2$^\textrm{nd}$ moment as a function of smoothing scale. We have also quantified how well the methods reconstruct the topology of the maps through the use of Minkowski functionals, and found that the noise properties of the forward fitted maps avoid introducing a bias present in the direct inversion which would influence the inference of cosmological constraints. \par

Our approach has been applied to the Dark Energy Survey Year 1 data and compared to the maps produced through direct inversion. This comparison finds broad agreement in central regions of the footprint, but disagreement at its edges. Our simulation tests indicate that these differences are symptomatic of errors introduced in the direct inversion at map edges, and that these areas of the sky are more reliably reconstructed in the forward fitted map. Our method provides a new opportunity to produce larger weak lensing mass maps to higher accuracy. Further work using this approach is worth pursuing, including producing maps to a higher resolution, use of priors on the fits, and comparison of cosmological constraints found using both the direct inversion and the fitted maps. 

\section{Acknowledgements}
BM would like to thank Gary Burton, Paul Carter, Robert Hardwick and Andrea Petri for useful discussion. DB is supported by STFC consolidated grant ST/S000550/1. Numerical computations were carried out on the Sciama High Performance Compute (HPC) cluster which is supported by the ICG, SEPNet and the University of Portsmouth. Some results in this paper have been derived using the HEALPix \citep{healpix_2005ApJ...622..759G} package. Figures were made using the \texttt{matplotlib} python plotting library \citep{Hunter:2007}.\par

Funding for the DES Projects has been provided by the U.S. Department of Energy, the U.S. National Science Foundation, the Ministry of Science and Education of Spain, 
the Science and Technology Facilities Council of the United Kingdom, the Higher Education Funding Council for England, the National Center for Supercomputing 
Applications at the University of Illinois at Urbana-Champaign, the Kavli Institute of Cosmological Physics at the University of Chicago, 
the Center for Cosmology and Astro-Particle Physics at the Ohio State University,
the Mitchell Institute for Fundamental Physics and Astronomy at Texas A\&M University, Financiadora de Estudos e Projetos, 
Funda{\c c}{\~a}o Carlos Chagas Filho de Amparo {\`a} Pesquisa do Estado do Rio de Janeiro, Conselho Nacional de Desenvolvimento Cient{\'i}fico e Tecnol{\'o}gico and 
the Minist{\'e}rio da Ci{\^e}ncia, Tecnologia e Inova{\c c}{\~a}o, the Deutsche Forschungsgemeinschaft and the Collaborating Institutions in the Dark Energy Survey. 

The Collaborating Institutions are Argonne National Laboratory, the University of California at Santa Cruz, the University of Cambridge, Centro de Investigaciones Energ{\'e}ticas, 
Medioambientales y Tecnol{\'o}gicas-Madrid, the University of Chicago, University College London, the DES-Brazil Consortium, the University of Edinburgh, 
the Eidgen{\"o}ssische Technische Hochschule (ETH) Z{\"u}rich, 
Fermi National Accelerator Laboratory, the University of Illinois at Urbana-Champaign, the Institut de Ci{\`e}ncies de l'Espai (IEEC/CSIC), 
the Institut de F{\'i}sica d'Altes Energies, Lawrence Berkeley National Laboratory, the Ludwig-Maximilians Universit{\"a}t M{\"u}nchen and the associated Excellence Cluster Universe, 
the University of Michigan, the National Optical Astronomy Observatory, the University of Nottingham, The Ohio State University, the University of Pennsylvania, the University of Portsmouth, 
SLAC National Accelerator Laboratory, Stanford University, the University of Sussex, Texas A\&M University, and the OzDES Membership Consortium.

Based in part on observations at Cerro Tololo Inter-American Observatory, National Optical Astronomy Observatory, which is operated by the Association of 
Universities for Research in Astronomy (AURA) under a cooperative agreement with the National Science Foundation.

The DES data management system is supported by the National Science Foundation under Grant Numbers AST-1138766 and AST-1536171.
The DES participants from Spanish institutions are partially supported by MINECO under grants AYA2015-71825, ESP2015-66861, FPA2015-68048, SEV-2016-0588, SEV-2016-0597, and MDM-2015-0509, 
some of which include ERDF funds from the European Union. IFAE is partially funded by the CERCA program of the Generalitat de Catalunya.
Research leading to these results has received funding from the European Research
Council under the European Union's Seventh Framework Program (FP7/2007-2013) including ERC grant agreements 240672, 291329, and 306478.
We  acknowledge support from the Brazilian Instituto Nacional de Ci\^encia
e Tecnologia (INCT) e-Universe (CNPq grant 465376/2014-2).

This manuscript has been authored by Fermi Research Alliance, LLC under Contract No. DE-AC02-07CH11359 with the U.S. Department of Energy, Office of Science, Office of High Energy Physics. The United States Government retains and the publisher, by accepting the article for publication, acknowledges that the United States Government retains a non-exclusive, paid-up, irrevocable, world-wide license to publish or reproduce the published form of this manuscript, or allow others to do so, for United States Government purposes.

\bibliographystyle{mnras}
\bibliography{FFWF}
\appendix
\section{Noise modelling}
\label{app:noise_modelling}
\begin{figure}
\includegraphics[width=0.5\textwidth]{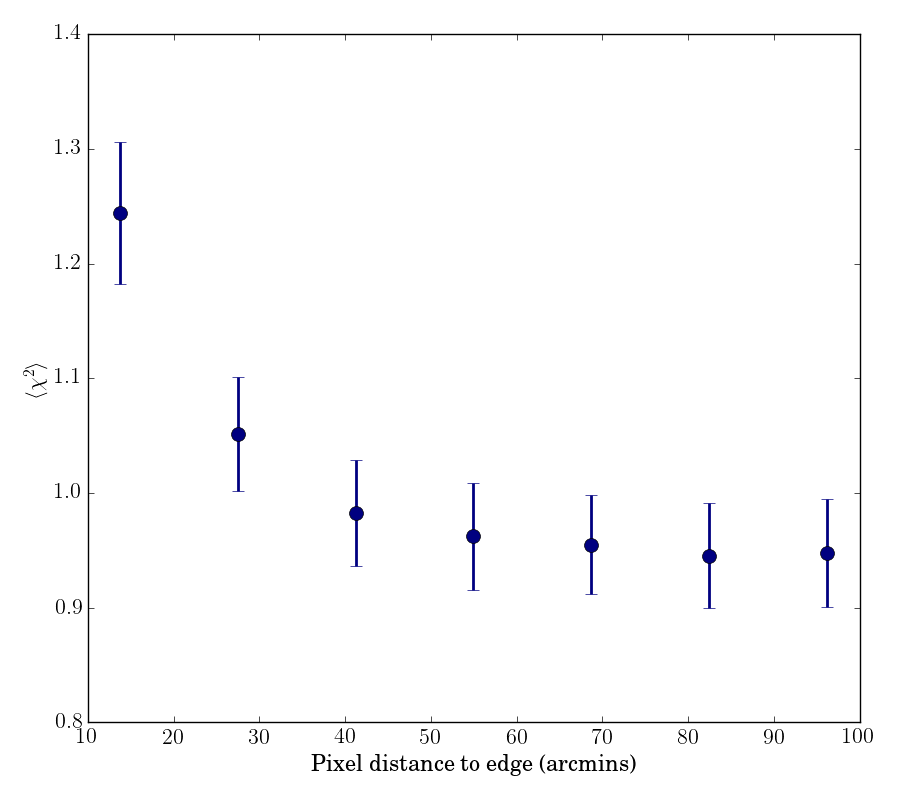}
\caption{The average $\chi^2$ for a set of pixels, as a function of the distance of these pixels to the edge of the fitted region, is shown for an ensemble of fits. This highlights how the routine preferentially constrains $\gamma$ values in the centre of the map at the expense of producing worse $\gamma$ estimates in the outer areas of the map. The overall fit to all pixels has $\chi^2/n = 1$, but as can be seen this does not force the variance for an outer pixel to be the same as that in the central region of the map. 
}
\label{fig:chi_edges}
\end{figure}
\begin{figure}
\label{fig:variance_edges}
\end{figure}
\begin{figure}
\label{fig:error_var_est}
\end{figure}
\begin{figure*}
\includegraphics[width=\textwidth]{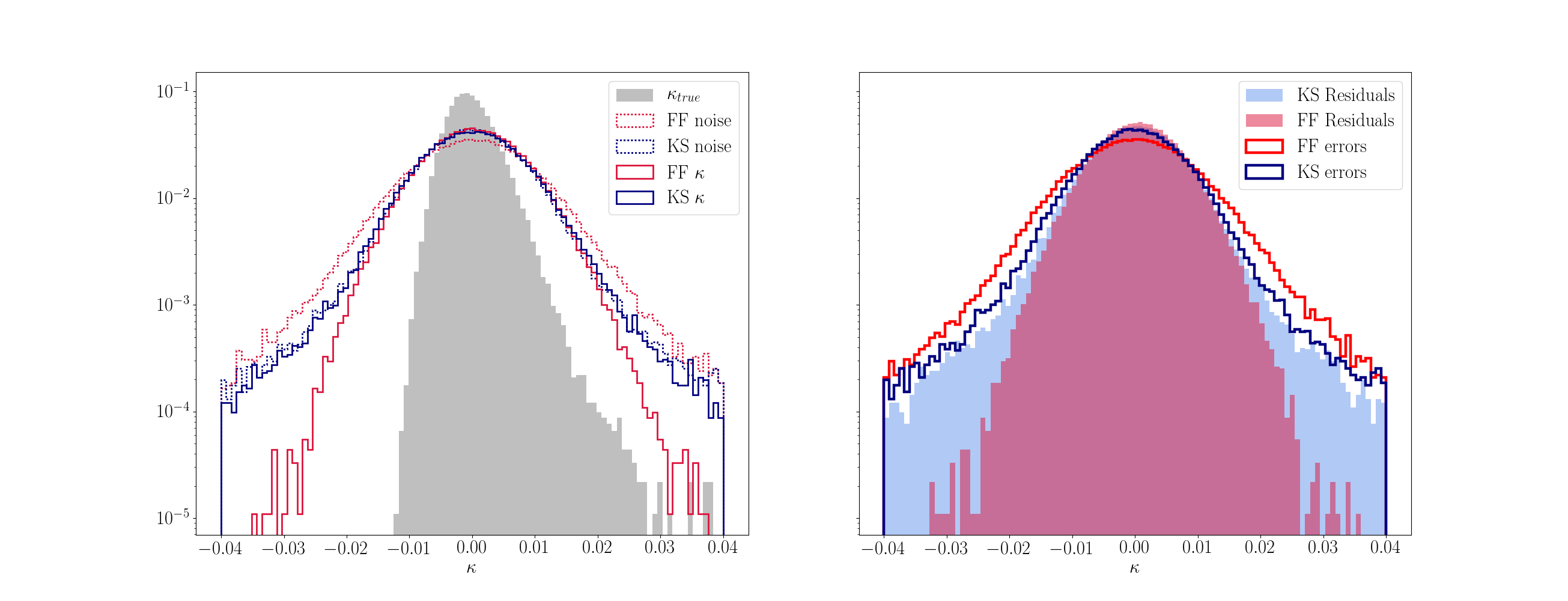}
\caption{Left: Distribution of convergence and associated noise for the Buzzard simulation and reconstruction. The solid lines represent reconstructed $\kappa$ distributions for Buzzard. The dashed lines show expected noise, if we estimate error in each pixel from the standard deviation of $\kappa$ values in an ensemble of reconstructions. Note that the dashed estimated noise distributions appear much wider than the solid reconstructed distributions (see text for explanation). 
Right: Distribution of noise estimated in two ways. The solid lines show the estimate from s.d. on $\kappa$ from an ensemble of reconstructions. The filled regions show residuals directly measured between our mean reconstruction and the true convergence field. }
\label{fig:edge_in_res}
\end{figure*}
Accurate quantification of the uncertainties for pixels in the fitted convergence maps is integral to understanding their reliability. 
 
A final fitted map is produced through averaging 40 maps resulting from our method pipeline. However, the error on the claimed convergence in each pixel of this final map is difficult to quantify. It might be thought that this could be well estimated from the standard deviation of the convergence across these 40 fits in a given pixel, but this is found, for edge pixels, to be much larger than the typical residuals between the final map and the true convergence. 

This can be understood by noting that a good overall fit to the data can be obtained even if the relatively small number of edge pixel values are a poorer fit. Figure \ref{fig:chi_edges} shows how the mean $\chi^2$ of the fitted shear changes as a function of the distance from the edge of the map; edge pixels are consistently further from $\chi^2/n_{\rm pix} =1$ in all of our fitted mapsF. This produces a larger spread of realisations of $\kappa$ in these pixels, and its standard deviation therefore has a larger range than the residuals between mean maps and the truth, as shown in Figure \ref{fig:edge_in_res}. 
\par

\section{Calculation of Minkowski functionals}
\label{app:functionals}
We follow the approach used in \citet{Hik_2006ApJ...653...11H} Appendix A to measure the Minkowski Functionals on the sphere. This approach considers the curvature of the sky and a pixelised surface. The three functionals for an excursion of size $\nu$ in a field $u$ normalised by its standard deviation, are described by
\begin{equation}
V_0=\mathcal{H}(u-\nu),
\end{equation}
\begin{equation}
V_1=\frac{1}{4}F(u-\nu)\sqrt[]{u_{;\theta}^2 + u_{;\phi}^2},
\end{equation}
\begin{equation}
V_2=\frac{1}{2\pi}F(u-\nu)\frac{2u_{;\theta}u_{;\phi}u_{;\theta \phi}- u_{;\theta}^2 u_{;\phi\phi}- u_{;\phi}^2 u_{;\theta \theta}}{u_{;\theta}^2 + u_{;\phi}^2}.
\end{equation}
The coordinates $\theta, \phi$ refer to angular positions on the sky and a semicolon denotes a partial derivative. We use this formalism in a \texttt{python} code to calculate the functionals from \texttt{HEALpix} maps. 
\section{Author affiliations}

$^{1}$ Institute of Cosmology and Gravitation, University of Portsmouth, Portsmouth, PO1 3FX, UK\\
$^{2}$ Department of Astronomy and Astrophysics, University of Chicago, Chicago, IL 60637, USA\\
$^{3}$ Kavli Institute for Cosmological Physics, University of Chicago, Chicago, IL 60637, USA\\
$^{4}$ Department of Astrophysical Sciences, Princeton University, Peyton Hall, Princeton, NJ 08544, USA\\
$^{5}$ Department of Physics, University of Arizona, Tucson, AZ 85721, USA\\
$^{6}$ Max Planck Institute for Extraterrestrial Physics, Giessenbachstrasse, 85748 Garching, Germany\\
$^{7}$ Universit\"ats-Sternwarte, Fakult\"at f\"ur Physik, Ludwig-Maximilians Universit\"at M\"unchen, Scheinerstr. 1, 81679 M\"unchen, Germany\\
$^{8}$ Department of Physics \& Astronomy, University College London, Gower Street, London, WC1E 6BT, UK\\
$^{9}$ Institut de F\'{\i}sica d'Altes Energies (IFAE), The Barcelona Institute of Science and Technology, Campus UAB, 08193 Bellaterra (Barcelona) Spain\\
$^{10}$ Institut d'Estudis Espacials de Catalunya (IEEC), 08034 Barcelona, Spain\\
$^{11}$ Institute of Space Sciences (ICE, CSIC),  Campus UAB, Carrer de Can Magrans, s/n,  08193 Barcelona, Spain\\
$^{12}$ Department of Physics, Stanford University, 382 Via Pueblo Mall, Stanford, CA 94305, USA\\
$^{13}$ Kavli Institute for Particle Astrophysics \& Cosmology, P. O. Box 2450, Stanford University, Stanford, CA 94305, USA\\
$^{14}$ SLAC National Accelerator Laboratory, Menlo Park, CA 94025, USA\\
$^{15}$ Department of Physics, ETH Zurich, Wolfgang-Pauli-Strasse 16, CH-8093 Zurich, Switzerland\\
$^{16}$ Department of Physics, Carnegie Mellon University, Pittsburgh, Pennsylvania 15312, USA\\
$^{17}$ Brookhaven National Laboratory, Bldg 510, Upton, NY 11973, USA\\
$^{18}$ Department of Physics, Duke University Durham, NC 27708, USA\\
$^{19}$ Institute for Astronomy, University of Edinburgh, Edinburgh EH9 3HJ, UK\\
$^{20}$ Cerro Tololo Inter-American Observatory, National Optical Astronomy Observatory, Casilla 603, La Serena, Chile\\
$^{21}$ Fermi National Accelerator Laboratory, P. O. Box 500, Batavia, IL 60510, USA\\
$^{22}$ CNRS, UMR 7095, Institut d'Astrophysique de Paris, F-75014, Paris, France\\
$^{23}$ Sorbonne Universit\'es, UPMC Univ Paris 06, UMR 7095, Institut d'Astrophysique de Paris, F-75014, Paris, France\\
$^{24}$ Jodrell Bank Center for Astrophysics, School of Physics and Astronomy, University of Manchester, Oxford Road, Manchester, M13 9PL, UK\\
$^{25}$ Centro de Investigaciones Energ\'eticas, Medioambientales y Tecnol\'ogicas (CIEMAT), Madrid, Spain\\
$^{26}$ Laborat\'orio Interinstitucional de e-Astronomia - LIneA, Rua Gal. Jos\'e Cristino 77, Rio de Janeiro, RJ - 20921-400, Brazil\\
$^{27}$ Department of Astronomy, University of Illinois at Urbana-Champaign, 1002 W. Green Street, Urbana, IL 61801, USA\\
$^{28}$ National Center for Supercomputing Applications, 1205 West Clark St., Urbana, IL 61801, USA\\
$^{29}$ Observat\'orio Nacional, Rua Gal. Jos\'e Cristino 77, Rio de Janeiro, RJ - 20921-400, Brazil\\
$^{30}$ Department of Physics, IIT Hyderabad, Kandi, Telangana 502285, India\\
$^{31}$ Department of Astronomy, University of Michigan, Ann Arbor, MI 48109, USA\\
$^{32}$ Department of Physics, University of Michigan, Ann Arbor, MI 48109, USA\\
$^{33}$ Instituto de Fisica Teorica UAM/CSIC, Universidad Autonoma de Madrid, 28049 Madrid, Spain\\
$^{34}$ Santa Cruz Institute for Particle Physics, Santa Cruz, CA 95064, USA\\
$^{35}$ Center for Cosmology and Astro-Particle Physics, The Ohio State University, Columbus, OH 43210, USA\\
$^{36}$ Department of Physics, The Ohio State University, Columbus, OH 43210, USA\\
$^{37}$ Center for Astrophysics $\vert$ Harvard \& Smithsonian, 60 Garden Street, Cambridge, MA 02138, USA\\
$^{38}$ Department of Physics and Astronomy, University of Pennsylvania, Philadelphia, PA 19104, USA\\
$^{39}$ Australian Astronomical Optics, Macquarie University, North Ryde, NSW 2113, Australia\\
$^{40}$ Departamento de F\'isica Matem\'atica, Instituto de F\'isica, Universidade de S\~ao Paulo, CP 66318, S\~ao Paulo, SP, 05314-970, Brazil\\
$^{41}$ George P. and Cynthia Woods Mitchell Institute for Fundamental Physics and Astronomy, and Department of Physics and Astronomy, Texas A\&M University, College Station, TX 77843,  USA\\
$^{42}$ Instituci\'o Catalana de Recerca i Estudis Avan\c{c}ats, E-08010 Barcelona, Spain\\
$^{43}$ School of Physics and Astronomy, University of Southampton,  Southampton, SO17 1BJ, UK\\
$^{44}$ Instituto de F\'isica Gleb Wataghin, Universidade Estadual de Campinas, 13083-859, Campinas, SP, Brazil\\
$^{45}$ Computer Science and Mathematics Division, Oak Ridge National Laboratory, Oak Ridge, TN 37831\\
$^{46}$ Argonne National Laboratory, 9700 South Cass Avenue, Lemont, IL 60439, USA\\

\end{document}